\documentclass[graybox, nosecnum]{svmult}

\usepackage{mathptmx}       
\usepackage{helvet}         
\usepackage{courier}        
\usepackage{type1cm}        
\usepackage{makeidx}         
\usepackage{graphicx}        
\usepackage{multicol}        
\usepackage[bottom]{footmisc}
\usepackage{hyperref}        
\usepackage{soul}            
\usepackage{wrapfig}
\usepackage{sidecap}
\makeindex             

\newcommand{\be}{\begin{equation}}
\newcommand{\ee}{\end{equation}}
\newcommand{\bea}{\begin{eqnarray}}
\newcommand{\eea}{\end{eqnarray}}
\begin{document}
\title*{Model for collective vibration}
\author{Haozhao Liang 
and Elena Litvinova}
\institute{Haozhao Liang \at Department of Physics, Graduate School of Science, The University of Tokyo, Tokyo 113-0033, Japan \email{haozhao.liang@phys.s.u-tokyo.ac.jp}
\and Elena Litvinova \at Department of Physics, Western Michigan University, Kalamazoo, MI 49008;
National Superconducting Cyclotron Laboratory, Michigan State University, East Lansing, MI 48824, USA
 \email{elena.litvinova@wmich.edu}}
%
%
\maketitle
\abstract{
We review the theory of nuclear collective vibrations evolved over decades from phenomenological quasiclassical picture to sophisticated microscopic approaches.
The major focus is put on the underlying microscopic mechanisms of emergent effects, which define the properties of giant resonances and soft modes.
The response of atomic nuclei to electromagnetic and weak fields is discussed in detail.
Astrophysical implications of the giant resonances and soft modes are outlined.
}

\section{\textit{Introduction}}

First ideas about the collective motion in atomic nuclei are associated with the name of Arkady Migdal and his seminal article ``Quadrupole and dipole $\gamma$-emission from nuclei'' \cite{Migdal1945}. Based on a semiclassical analysis of the induced nuclear dipole moment in a uniform electric field and on the electric dipole sum rule, a relation between the average energy of the electric dipole transitions and the symmetry term of the  Weizs\"acker's binding energy formula was established. Independently of Migdal, Goldhaber and Teller \cite{Goldhaber1948} obtained expressions for the resonance energy of classical harmonic oscillations of protons with respect to neutrons, while analyzing the experimental data of Ref.~\cite{BaldwinKlaiber1947}
on the photoabsorption and  photofission, where a broad peak in the cross sections with the centroid above $10$~MeV was observed. Since these pivotal works, the collective oscillation of the proton and neutron Fermi liquids against each other is associated with the dipole excitations and acquired the name of giant dipole resonance (GDR), as it was found dominant in the observed 
photoabsorption spectra of all nuclei. Because of its very pronounced character, the GDR dominated the studies of nuclear collective excitations for decades, see a recent comprehensive review in Ref.~\cite{Ishkhanov2021}. During this time, other giant resonances, including the charge-neutral giant monopole resonance (GMR) and giant quadrupole resonance (GQR) as well as the charge-exchange (also called spin-isospin) isobaric analog state (IAS), Gamow-Teller resonance (GTR), and spin-dipole resonance (SDR), have also been discovered and investigated \cite{Osterfeld1992, GR2001, Ichimura2006, PaarVretenarKhanEtAl2007, Roca-Maza2018}.


A microscopic interpretation of the GDR has become possible within the shell model concept introduced by Mayer and Jensen \cite{MayerJensen1955}, which was realized, in particular, by Wilkinson in Ref.~\cite{Wilkinson1956}.  In the single-particle shell model, the electric dipole excitations are formed by the one-particle-one-hole ($1p1h$) transitions from the occupied shell to the nearest unoccupied shell. In this picture, the GDR centroid should be close to the mean separation energy between the shells, while the transition energies group within a relatively narrow energy interval. However, this model disagrees with the collective nature of the GDR and, moreover, strongly underestimates its energy observed in experiment. 


The next major step in the theoretical description of the GDR was made with the understanding that the residual interaction between the particles and holes has to be included in the theory \cite{EliotFlowers1957,BrownBolsterli1959}. The random phase approximation (RPA), originally developed for electronic systems \cite{BohmPines1951}, became very successful in resolving the conflict between the single-particle and collective features of the GDR. Since then, RPA was employed as a major theoretical tool for investigating nuclear collective excitations, because of the relative simplicity of its general equations \cite{RingSchuck1980}. Formulated in the mean-field basis and invoking the notion of the effective residual interaction, RPA is compatible with various implementation algorithms. It was, thus, widely applied to the description of collective excitations of various multipolarities \cite{PaarVretenarKhanEtAl2007, Roca-Maza2018}, which were actively studied experimentally, too \cite{Osterfeld1992, GR2001, Ichimura2006}.



Thus, RPA was recognized quite early as a good approach to the gross features of nuclear spectra, such as the locations of the main peaks and the sum rules of collective excitations \cite{Migdal1967}. However, the quantitative description of the giant resonance's width formation, which originates from the damping effects, required extensions beyond RPA.
Bohr and Mottelson \cite{BohrMottelson1969,BohrMottelson1975} came with the idea that the spectra can be further refined, if the coupling between the single-particle and emergent collective degrees of freedom is taken into account. Since then, various realizations of this idea, among which the nuclear field theory (NFT) \cite{BertschBortignonBroglia1983} and the quasiparticle-phonon model (QPM) \cite{Soloviev1992} were the most prominent ones, explained successfully many of the observed phenomena. 
Although based on the effective Hamiltonians of quasiparticles and phonons with phenomenological interactions, these concepts can be, 
in fact, linked to the non-perturbative versions of quantum-field-theory-based equations of motion (EOM) for the fermionic correlation functions in nuclear medium \cite{RingSchuck1980}. The EOM method developed and further elaborated, e.g., in Refs.~\cite{Rowe1968,Schuck1976}, being, in principle, exact, allows for a hierarchy of approximations to the dynamical kernels of the equations for the one-fermion and two-time few-fermion propagators. The full resummations in the particle-hole ($ph$) and particle-particle ($pp$) channels, which are implied in the non-perturbative versions of those kernels truncated at the two-body level, can be mapped to the kernels of NFT and QPM. 


In the latter theories, such kernels originate from the particle-vibration coupling (PVC)  \cite{BohrMottelson1969,BohrMottelson1975,BertschBortignonBroglia1983,KamerdzhievTertychnyiTselyaev1997}
or multiphonon \cite{Soloviev1992,Ponomarev1999b,SavranBabilonBergEtAl2006,LoIudice2012} configurations, which are associated with the emergent collective phenomena in the strongly-coupled regimes.   
The above-mentioned mapping to the fermionic EOM's, addressed, in particular, in Ref.~\cite{LitvinovaSchuck2019}, has to be corrected for the accurate lowest-order limit \cite{RingSchuck1980}, however, such corrections are found to be small for nuclear systems \cite{giai1983}.
Unfortunately, in the past decades, the computational capabilities did not allow for the calculations within large and complex model spaces as well as for keeping direct connections to the underlying bare nucleon-nucleon forces in the interaction kernels.  Nevertheless, the semi-phenomenological PVC and QPM models based on the effective in-medium interactions or G-matrix theory  \cite{BertschBortignonBroglia1983,KamerdzhievTertychnyiTselyaev1997,Ponomarev1999b,LoIudice2012,MahauxBortignonBrogliaEtAl1985,Tselyaev1989,Pankratov:2011hi} provided over the years the important knowledge about the coupling between the single-particle and collective degrees of freedom in nuclei. Later, such approaches were linked to the contemporary density functional theories (DFT) \cite{LitvinovaRing2006,LitvinovaRingTselyaev2008,LitvinovaRingTselyaev2010,LitvinovaWibowo2018,Tselyaev2016,NiuNiuColoEtAl2015,Saperstein2016,
Shen2020}, advancing the PVC models to the self-consistent frameworks. Recently, a few attempts of combining the multiphonon approaches with the so-called optimized chiral interactions were presented \cite{DeGregorio2017,Knapp:2014xja}. This method demonstrated some promising results for light nuclear systems, however, a satisfactory description of heavy nuclei was only possible by introducing an additional three-body interaction with a free strength parameter.

Under the assumption of weak coupling, in the leading approximation, the dynamical kernels are composed of the products of one-fermion propagators and fully neglect the higher-rank ones. Although this assumption is not fully justified for nuclear systems, such approaches to the nuclear responses as the second RPA \cite{Yannouleas1983} show reasonably good results for nuclear excitations in various channels. For instance, Ref.~\cite{Drozdz:1990zz} presented a comprehensive review on the nuclear spin-isospin responses in medium-mass nuclei with the G-matrix interaction and showed a systematic improvement of the description of this type of responses, as compared to RPA. More recent implementations of the second RPA with the effective interactions of Skyrme type devoted mostly to the neutral giant resonances \cite{Grasso:2020uik}
and demonstrated some progress after applying the subtraction technique proposed originally in Ref.~\cite{Tselyaev2013}. Very interesting attempts of the second RPA calculations for nuclear excited states investigated the potential of this approach to advance the \textit{ab-initio} theories to medium-mass nuclei  \cite{PapakonstantinouRoth2009}. Developments in a similar direction based on modern interactions showed a substantial progress on the single-particle dynamical kernels with perturbative treatments of various kinds  \cite{Soma:2012zd,Cipollone:2013zma,Lu:2017nbi} and on analogous concepts for nuclear response to the electromagnetic probes \cite{Bacca:2013dma,Raimondi:2018mtv}.


\section{\textit{The exact equation of motion for nuclear response and approximate methods}}

The conventional ways of deriving the equations describing nuclear collective vibrations can be found in many 
textbooks, however, they are mostly confined by the RPA equations, either in the matrix or in the linear response forms, see, for instance, Ref.~\cite{RingSchuck1980}. A  remarkably efficient finite amplitude method for solving the RPA equations numerically,
without explicit calculations of the interaction matrix elements,  has been developed in Ref. \cite{Nakatsukasa2007} and attracted intensive attention during the past decade.
Going beyond the static-kernel approximation, one can extend the RPA to higher-order RPAs, the (quasi)particle-vibration coupling and quasiparticle-phonon approaches by, for instance, supplementing the traditional RPA $1p1h$ excitation operator with the terms of growing complexity.

In contrast, in this Chapter the authors derive the equations for nuclear collective vibrations in a top-down way. They start from the exact equation of motion for nuclear response, which allows for a hierarchy of approximations, including the special cases of the second RPA, PVC, and QPM, as well as the conventional RPA.

The nuclear response to an external field of one-body character is completely determined by the response function
\be
R(12,1'2') \equiv R_{12,1'2'}(t-t') =  -i\langle T\psi^{\dagger}(1)\psi(2)\psi^{\dagger}(2')\psi(1')\rangle,
\label{phresp}
\ee
which is a ground-state average of the time-ordered fermionic field operators $\psi$ and $\psi^{\dagger}$ in the Heisenberg picture, such as
\be
\psi(1) \equiv \psi_1(t_1) \equiv {e}^{iHt_1}\psi_1 {e}^{-iHt_1},
\qquad\qquad
\psi^{\dagger}(1) \equiv \psi^{\dagger}_1(t_1) \equiv {e}^{iHt_1}\psi^{\dagger}_1 {e}^{-iHt_1}.
\label{t-fields}
\ee 
In Eqs.~(\ref{phresp}) and (\ref{t-fields}), the number subscripts denote the complete sets of single-particle quantum numbers, and it is implied in Eq.~(\ref{phresp}) that $t_1 = t_2 = t$ and $t_{1'} = t_{2'} = t'$, i.e., a two-time particle-hole correlation function is considered. Furthermore, the convention $\hbar = 1$ is adopted throughout this Chapter.  The operator $H$ in Eq.~(\ref{t-fields}) is the standard many-body fermionic Hamiltonian,
\be
H = H^{(1)} + V^{(2)},
\label{Hamiltonian}
\ee
confined, for definiteness, by the two-body interaction $V^{(2)}$.
The one-body term
\be
H^{(1)} = \sum_{12} t_{12} \psi^{\dag}_1\psi_2 + \sum_{12}v^{(\rm MF)}_{12}\psi^{\dag}_1\psi_2 \equiv \sum_{12}h_{12}\psi^{\dag}_1\psi_2
\label{Hamiltonian1}
\ee
is defined by the matrix elements $h_{12}$ which combine, in general, the kinetic energy $t$ and the mean-field $v^{(\rm MF)}$ part of the interaction. The two-body interaction is described by the operator
\be
V^{(2)} = \frac{1}{4}\sum\limits_{1234}{\bar v}_{1234}{\psi^{\dagger}}_1{\psi^{\dagger}}_2\psi_4\psi_3,
\label{Hamiltonian2}
\ee
where ${\bar v}_{1234} = v_{1234} - v_{1243}$ is the antisymmetrized matrix element of the  interaction of two fermions in the vacuum. In this Section, we will confine the formalism by the non-relativistic Hamiltonians, however, it can be straightforwardly generalized to the relativistic ones, both bare and effective \cite{Walecka1974,  VretenarAfanasjevLalazissisEtAl2005, Meng2006, Liang2015, Meng2016}. By the related reasons, the authors omit the details associated with the origin of the two-body interaction $V^{(2)}$ assuming, however, its instantaneous, or time-independent, character. Furthermore, the framework can also be directly extended by the three-body forces, which are omitted here as the majority of implementations are confined by the two-body Hamiltonians.

The Fourier image of the response function (\ref{phresp}) in the energy (frequency) domain reads
\be
R_{12,1'2'}(\omega) = \sum\limits_{\nu>0}\Bigl[ \frac{\rho^{\nu}_{21}\rho^{\nu\ast}_{2'1'}}{\omega - \omega_{\nu} + i\delta} -  \frac{\rho^{\nu\ast}_{12}\rho^{\nu}_{1'2'}}{\omega + \omega_{\nu} - i\delta}\Bigr],
\label{respspec}
\ee
being the spectral expansion over the exact excited states $|\nu\rangle$ of the many-body system with the energies $\omega_{\nu} = E_{\nu} - E_0$ measured from the ground-state energy. The residues of this expansion are the products of the transition densities
\be
\rho^{\nu}_{12} = \langle 0|\psi^{\dagger}_2\psi_1|\nu \rangle , 
\label{trden}
\ee
which represent the weights of the pure particle-hole configurations in the single-particle basis $\{1\}$ on top of the ground state $|0\rangle$ in the exact excited states $|\nu\rangle$. 

The EOM for the response function can be generated by the differentiation of Eq.~(\ref{phresp}) with respect to the time arguments. Differentiation with respect to $t$ leads to
\be
(i\partial_t + \varepsilon_{12})R_{12,1'2'}(t-t')
= \delta(t-t'){\cal N}_{121'2'} 
+ i\langle T[V,{\psi^{\dagger}}_1\psi_2](t)({\psi^{\dagger}}_{2'}\psi_{1'})(t')\rangle ,
\label{dtG2b}                           
\ee
where the norm kernel is introduced as
\be
{\cal N}_{121'2'} = \langle[\psi^{\dagger}_{1}\psi_{2},\psi^{\dagger}_{2'}\psi_{1'}]\rangle =  \delta_{22'}\langle \psi^{\dagger}_{1}\psi_{1'} \rangle - 
\delta_{11'}\langle \psi^{\dagger}_{2'}\psi_{2} \rangle. 
\label{norm}
\ee
With the diagonal one-body density matrix, the norm simplifies to the form of ${\cal N}_{121'2'} = \delta_{11'}\delta_{22'}(n_1 - n_2) \equiv \delta_{11'}\delta_{22'}{\cal N}_{12}$, where
$n_1 = \langle{\psi^{\dagger}}_1\psi_1\rangle$ is identified with the occupancy of the fermionic state $1$. In Eq.~(\ref{dtG2b}) and in the following, $\varepsilon_{12} = \varepsilon_{1} - \varepsilon_{2}$ with $\varepsilon_{1}$ and $\varepsilon_{2}$ being the eigenvalues of the one-body part of the Hamiltonian~(\ref{Hamiltonian1}).

The differentiation of the last term on the right hand side of Eq.~(\ref{dtG2b}) with respect to $t'$ generates the second EOM,
\bea
i\langle T[V,{\psi^{\dagger}}_1\psi_2](t)({\psi^{\dagger}}_{2'}\psi_{1'})(t')\rangle(-i\overleftarrow{\partial_{t'}} &-& \varepsilon_{2'1'}) 
= -\delta(t-t')
\langle [[V,{\psi^{\dagger}}_1\psi_2],{\psi^{\dagger}}_{2'}\psi_{1'}]\rangle \nonumber \\ &+& i\langle T[V,{\psi^{\dagger}}_1\psi_2](t)[V,{\psi^{\dagger}}_{2'}\psi_{1'}](t')\rangle.
\label{dtG2c}
\eea
Combining Eqs.~(\ref{dtG2c}) and (\ref{dtG2b}), after the Fourier transformation to the energy domain, one obtains 
\be
R_{12,1'2'}(\omega) = R^{(0)}_{12,1'2'}(\omega) 
+ \sum\limits_{343'4'} R^{(0)}_{12,34}(\omega)T_{34,3'4'}(\omega)R^{(0)}_{3'4',1'2'}(\omega),
\label{2bgfb}
\ee
with the uncorrelated particle-hole response $R^{(0)}(\omega)$,
\be
R^{(0)}_{12,1'2'}(\omega) = \frac{{\cal N}_{121'2'}}{\omega - \varepsilon_{21}} = \delta_{11'}\delta_{22'}\frac{n_1 - n_2}{\omega - \varepsilon_{21}},
\label{resp0}
\ee
and the $T$-matrix $T(\omega)$, which is the Fourier image of
\bea
T_{12,1'2'}(t-t') &=& {\cal N}_{12}^{-1}\Bigl[-\delta(t-t')
\langle [[V,{\psi^{\dagger}}_1\psi_2],{\psi^{\dagger}}_{2'}\psi_{1'}]\rangle \nonumber \\
&~&+ i\langle T[V,{\psi^{\dagger}}_1\psi_2](t)[V,{\psi^{\dagger}}_{2'}\psi_{1'}](t')\rangle \Bigr] {\cal N}_{1'2'}^{-1}. 
\label{F1}
\eea
Thereby, the $T$-matrix is decomposed naturally into the instantaneous (static) $T^{(0)}$ and time-dependent (dynamical) $T^{(r)}$ terms,
\be\label{Ft} 
T_{12,1'2'}(t-t') = {\tilde{\cal N}}_{121'2'}^{-1}\Bigl(T^{(0)}_{12,1'2'}\delta(t-t') + T^{(r)}_{12,1'2'} (t-t')\Bigr),
\ee
with ${\tilde{\cal N}}_{121'2'} = {\cal N}_{12}{\cal N}_{1'2'}$ and
\bea
T^{(0)}_{12,1'2'} &=& -\langle [[V,{\psi^{\dagger}}_1\psi_2],{\psi^{\dagger}}_{2'}\psi_{1'}]\rangle, \nonumber \\
T^{(r)}_{12,1'2'}(t-t') &=&  i\langle T[V,{\psi^{\dagger}}_1\psi_2](t)[V,{\psi^{\dagger}}_{2'}\psi_{1'}](t')\rangle,
\label{Ft2} 
\eea
where the superscript "(r)" indicates the retarded character of the dynamical term.
It is convenient to further transform Eq.~(\ref{2bgfb}) into a formally closed equation for $R(\omega)$, similar to the Dyson equation for the one-fermion 
propagators, by introducing the kernel $K(\omega)$ irreducible with respect to the uncorrelated particle-hole response $R^{(0)}$, i.e.,
\be
R(\omega) = R^{(0)}(\omega) + R^{(0)}(\omega)K(\omega)R(\omega),
\label{Dyson2}
\ee
where
\be
T(\omega) = K(\omega) + K(\omega)R^{(0)}(\omega)T(\omega). 
\label{Kkernel}
\ee
In other words, $K(\omega) = T^{irr}(\omega)$ and it can also be decomposed into the instantaneous and time-dependent parts as
\be
K(t-t') = {\tilde{\cal N}}^{-1}\Bigl(K^{(0)}\delta(t-t') + K^{(r)}(t-t')\Bigr), 
\label{Womega}
\ee
with
\be
K^{(0)} = T^{(0)},
\qquad \qquad
K^{(r)}(t-t') = T^{(r)irr}(t-t').
\ee
In a complete analogy to the case of one-fermion EOM, the decomposition of the interaction kernel in Eqs.~(\ref{Ft}) and (\ref{Womega}) into the static and time-dependent, or dynamical, parts is a generic feature  of the in-medium interaction in the particle-hole channel and the direct consequence of the time-independence of the bare interaction $V$ of Eq.~(\ref{Hamiltonian2}). 

\begin{figure}[t]
\begin{center}
\includegraphics[width = 0.8\textwidth]{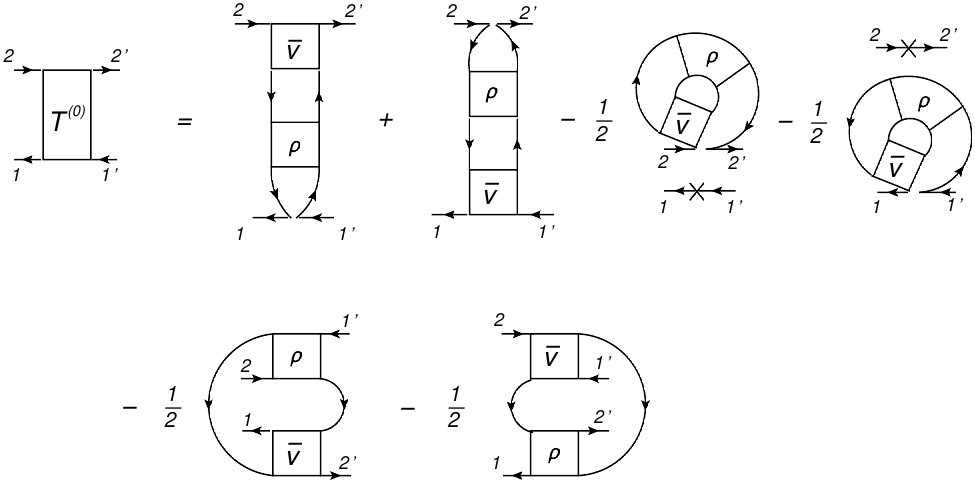}
\end{center}
\caption{Diagrammatic representation of the static part of the kernel $T^{(0)}_{12,1'2'}$ defined by Eqs.~(\ref{Ft2}) and (\ref{Fstatic1}). The lines with arrows denote fermionic propagators, while the rectangular blocks stand for the antisymmetrized nucleon-nucleon interaction $\bar v$ and the two-body density $\rho$ of Eq.~(\ref{rho2}). The figure is adopted from Ref. \cite{LitvinovaSchuck2019}.}
\label{SE2static}%
\end{figure}

After evaluating the commutators in Eqs.~(\ref{Ft2}) and introducing the two-fermion density 
\be
\rho_{ij,kl} = \langle\psi^{\dagger}_{k}\psi^{\dagger}_{l}\psi_{j}\psi_{i}\rangle = \rho_{ik}\rho_{jl} - \rho_{il}\rho_{jk} + \sigma^{(2)}_{ij,kl},
\label{rho2}
\ee
where $\sigma^{(2)}_{ij,kl}$ represents its fully correlated part and the Roman indices have the same meaning as the number indices, 
the static kernel takes the form of
\bea
K^{(0)}_{12,1'2'} &=&  {\cal N}_{12}{\bar v}_{21'12'}{\cal N}_{1'2'} 
+ \sum\limits_{jk}{\bar v}_{2j2'k} \sigma^{(2)}_{1'k,1j} + \sum\limits_{jk}
{\bar v}_{1'k1j}\sigma^{(2)}_{2j,2'k} \nonumber  \\
&& - \frac{1}{2}\delta_{11'}\sum\limits_{jkl}{\bar v}_{2jkl}\sigma^{(2)}_{kl,2'j} 
- \frac{1}{2}\delta_{22'}\sum\limits_{ijk}{\bar v}_{ji1k}\sigma^{(2)}_{1'k,ji} \nonumber  \\
&& - \frac{1}{2}\sum\limits_{ij}{\bar v}_{ij2'1}\sigma^{(2)}_{1'2,ij}
- \frac{1}{2}\sum\limits_{kl}{\bar v}_{21'kl}\sigma^{(2)}_{kl,12'}.
\label{Fstatic1}
\eea
In this form, the first term isolates the contribution from the bare interaction, where the norm factors are compensated by their inverses in Eq.~(\ref{Ft}). The remaining terms of $T^0$ with the single-particle mean field are absorbed in the single-particle energies by the substitution $\varepsilon_1 \to {\tilde{\varepsilon}}_1 = \varepsilon_1 + \Sigma^{(0)}_{11}$, where $\Sigma^{(0)}_{11} = \sum_{ij}{\bar v}_{1i1j}\langle \psi^{\dagger}_i\psi_j\rangle$, in the uncorrelated response of Eq.~(\ref{resp0}). 
Thereby, Eq.~(\ref{Fstatic1}) gives the exact form of the static kernel which, 
in the absence of correlations contained in the quantities $\sigma^{(2)}$ and $T^{(r)}$, reduces the EOM (\ref{Dyson2}) to the well-known RPA.
The static part of the in-medium two-fermion interaction kernel in the particle-hole channel is shown diagrammatically in Fig.~\ref{SE2static}.

\begin{figure}[t]
\begin{center}
\includegraphics[width = 0.65\textwidth]{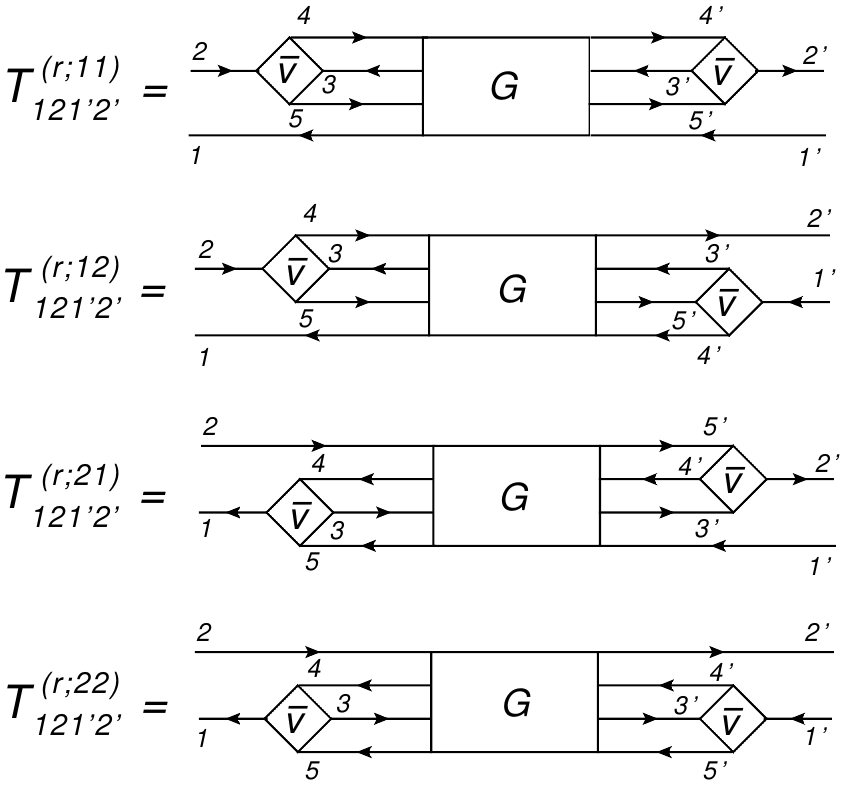}
\end{center}
\caption{Diagrammatic representation of the four components of the dynamical kernel $T^{(r)}_{12,1'2'}(t-t')$  of Eqs.~(\ref{Tr11})--(\ref{Tr22}). The blocks $G$ are associated with the corresponding time-ordered operator products.}
\label{SE2irrtot}%
\end{figure}

The dynamical part $T^{(r)}$ of the kernel $T$, after evaluating  the commutators of Eqs.~(\ref{Ft2}), is decomposed as
\be
T^{(r)}_{12,1'2'}(t-t') 
= T^{(r;11)}_{12,1'2'}(t-t') + T^{(r;12)}_{12,1'2'}(t-t') +T^{(r;21)}_{12,1'2'}(t-t') + T^{(r;22)}_{12,1'2'}(t-t'),\nonumber
\label{Tr_comp}
 \ee
where
\bea
\label{Tr11}
T^{(r;11)}_{12,1'2'}(t-t') &=&  
 -\frac{i}{4}\sum\limits_{jkl}{\bar v}_{j2kl}\langle T(\psi^{\dagger}_1\psi^{\dagger}_j\psi_l\psi_k)(t)
\sum\limits_{mnp}(\psi^{\dagger}_m\psi^{\dagger}_n\psi_p\psi_{1'})(t')\rangle {\bar v}_{nm2'p},\ \  \ \ \\
\label{Tr12}
T^{(r;12)}_{12,1'2'}(t-t') &=&   \frac{i}{4}\sum\limits_{jkl}
{\bar v}_{j2kl}\langle T (\psi^{\dagger}_1\psi^{\dagger}_j\psi_l\psi_k)(t)\sum\limits_{npq}(\psi^{\dagger}_{2'}\psi^{\dagger}_n\psi_q\psi_p)(t')\rangle{\bar v}_{n1'pq}, \\
\label{Tr21}
T^{(r;21)}_{12,1'2'}(t-t') &=&  
\frac{i}{4}\sum\limits_{ijk}{\bar v}_{ji1k} \langle T(\psi^{\dagger}_i\psi^{\dagger}_j\psi_k\psi_2)(t)
\sum\limits_{mnp}(\psi^{\dagger}_m\psi^{\dagger}_n\psi_p\psi_{1'})(t')\rangle {\bar v}_{nm2'p}, \\
\label{Tr22}
T^{(r;22)}_{12,1'2'}(t-t') &=&   
-\frac{i}{4}\sum\limits_{ijk}{\bar v}_{ji1k} \langle T(\psi^{\dagger}_i\psi^{\dagger}_j\psi_k\psi_2)(t)
\sum\limits_{npq}(\psi^{\dagger}_{2'}\psi^{\dagger}_n\psi_q\psi_p)(t')\rangle {\bar v}_{n1'pq},
\eea
and shown diagrammatically in Fig.~\ref{SE2irrtot}.  The operator products in Eqs.~(\ref{Tr11})--(\ref{Tr22}) define the correlated $2p2h$ propagators
\be
G(543'1',5'4'31) = \langle T(\psi^{\dagger}_1\psi^{\dagger}_3\psi_5\psi_4)(t)(\psi^{\dagger}_{4'}\psi^{\dagger}_{5'}\psi_{3'}\psi_{1'})(t')\rangle.
\label{2p2hGF}
\ee

Thus, the particle-hole response function (\ref{phresp}) is the solution of the integral equation
\be
R_{12,1'2'}(\omega) = {\tilde R}_{12,1'2'}(\omega) 
+ \sum\limits_{343'4'} {\tilde R}_{12,34}(\omega)\Bigl( K^{(0)}_{34,3'4'} +K^{(r)}_{34,3'4'}(\omega)\Bigr)R_{3'4',1'2'}(\omega),
\label{BSE-Dyson}
\ee
where $K^{(r)}_{12,1'2'}(\omega) = T^{(r)irr}_{12,1'2'}(\omega)$ and the mean-field response ${\tilde R}_{12,1'2'}(\omega)$ is defined as
\be
{\tilde R}_{12,1'2'}(\omega) =  \delta_{11'}\delta_{22'}\frac{{\tilde n}_1 - {\tilde n}_2}{\omega - {\tilde\varepsilon}_{21}},
\ee
with the occupancies ${\tilde n}_i$ in the basis, which diagonalizes $\varepsilon_i$.
Because of the two-time nature of the response function, its EOM reduces to the equation with only one energy (frequency) variable in the energy domain. The kernel $K(\omega)$, which is split into the static and dynamical parts, contains all the in-medium nucleonic correlations, which are, in principle, completely determined by the bare interaction $\bar v$.

However, in practice, consistent calculations of both the static and dynamical kernels are technically very difficult, and the difficulties are of a conceptual character. An accurate computation of the static kernel requires the correlated two-body density, and the dynamical kernel is based on the correlated $2p2h$ propagator.  While the two-body density can be extracted from the response function in the static limit, the correlated $2p2h$ propagator, in principle, requires a solution of the respective EOM.

Alternatively, the $2p2h$ propagator can be approximated by various cluster expansions over lower-rank propagators. The solutions neglecting completely the dynamical kernel are the simplest approaches to the nuclear response, which correspond to the RPA. Typically the RPA is formulated in terms of the transition densities~(\ref{trden}). Indeed, substituting the spectral form of $R_{12,1'2'}(\omega)$ in Eq.~(\ref{respspec}) into Eq.~(\ref{BSE-Dyson}) and dropping the dynamical kernel $K^{(r)}$, in the vicinity of the pole $\omega \to \omega_{\nu}$, one obtains the equation
\be
\rho^{\nu}_{12} = \frac{{\tilde n}_2 - {\tilde n}_1}{\omega - {\tilde\varepsilon}_{12}}\sum\limits_{34}K^{(0)}_{21,43}\rho^{\nu}_{34}
\label{RPA}
\ee
for the transition density $\rho^{\nu}$. Identifying its matrix elements with the $X^{\nu}$ and $Y^{\nu}$ amplitudes as $X^{\nu}_{ph} = \rho^{\nu}_{ph}$ and 
$Y^{\nu}_{ph} = \rho^{\nu}_{hp}$, Eq.~(\ref{RPA}) can be brought to the conventional $2\times 2$ block matrix form \cite{RingSchuck1980}. Retaining the full correlated static kernel leads to the self-consistent RPA \cite{Schuck2021}, while neglecting the terms with two-body correlations $\sigma^{(2)}$ implies the ordinary RPA. In the latter case, for a reasonable description of the observed spectra, the bare interaction should be replaced by an effective interaction, either schematic or derived from the DFT. Such interactions are designed to take into account the neglected correlations in a static approximation by absorbing them in the parameters. This type of approaches comprises most of the modern applications of RPA.
The quasiparticle RPA (QRPA) \cite{RingSchuck1980} can be obtained analogously in the basis of the Bogoliubov quasiparticles \cite{Bogoliubov1958}. 

Another type of approaches goes beyond (Q)RPA by retaining also the dynamical kernel $K^{(r)}$. These approaches can be further classified by the approximations employed for the treatment of this kernel, 
which is
irreducible in the particle-hole channel. 
The possible cluster expansions of the generic $2p2h$ propagator (\ref{2p2hGF}) entering all the four terms (\ref{Tr11})--(\ref{Tr22}),  which satisfy this condition and do not involve the correlated propagators of more than two fermions, are the following. 

\begin{figure}[t]
\begin{center}
\includegraphics[width = 0.8\textwidth]{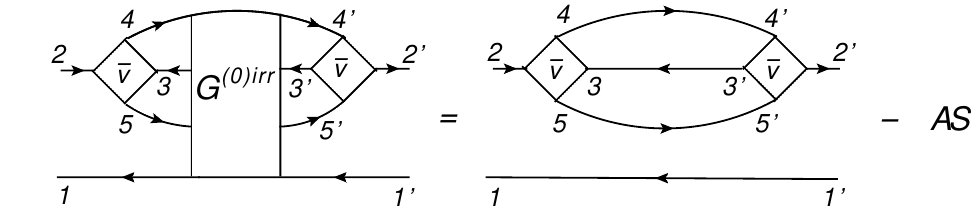}
\end{center}
\caption{Diagrammatic representation of the uncorrelated contributions (i) 
to the $(11)$-component of the dynamical kernel $T^{(r;11)irr}_{12,1'2'}(t-t')$ irreducible with respect to the particle-hole propagator. `$AS$' includes all the antisymmetrized contributions.}
\label{SE2irr0}%
\end{figure}

(i) The completely uncorrelated factorization reads
\bea
G^{(0)irr}(543'1',5'4'31)
 &=& \langle T({\psi^{\dagger}}_1{\psi^{\dagger}}_3)(t)({\psi}_{3'}{\psi}_{1'})(t')\rangle_0 \nonumber \\
&&\times \langle T({\psi}_5{\psi}_4)(t)({\psi^{\dagger}}_{4'}{\psi^{\dagger}}_{5'})(t')\rangle_0, 
\label{G0irr}
\eea
i.e., it is a product of two uncorrelated antisymmetrized propagators, such as
\bea
\langle T({\psi^{\dagger}}_1{\psi^{\dagger}}_3)(t)({\psi}_{3'}{\psi}_{1'})(t')\rangle_0 
&=& \langle T{\psi^{\dagger}}_1(t)\psi_{1'}(t')\rangle_0
\langle T{\psi^{\dagger}}_3(t)\psi_{3'}(t')\rangle_0 \nonumber\\
&& - \langle T{\psi^{\dagger}}_1(t)\psi_{3'}(t')\rangle_0
\langle T{\psi^{\dagger}}_3(t)\psi_{1'}(t')\rangle_0,
\eea
where the subscript `0' indicates the uncorrelated character of the quantity. This type of dynamical kernel is shown diagrammatically in Fig.~\ref{SE2irr0} for the term $T^{(r;11)irr}_{12,1'2'}(t-t')$ of Eq.~(\ref{Tr11}). It illustrates explicitly the $2p2h$ content of this approach,
which corresponds to various versions of the second RPA \cite{Yannouleas1983,Drozdz:1990zz,
Grasso:2020uik,PapakonstantinouRoth2009,Raimondi:2018mtv}. 

\begin{figure}[t]
\begin{center}
\includegraphics[width = 0.9\textwidth]{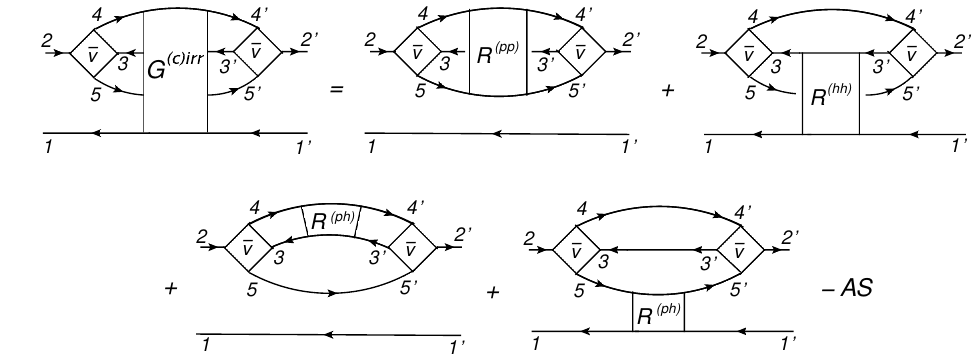}
\end{center}
\caption{Diagrammatic representation of the singly correlated approximation (ii) 
to the $(11)$-component of the dynamical kernel $T^{(r;11)irr}_{12,1'2'}(t-t')$ irreducible with respect to the particle-hole propagator. The rectangular blocks 
$R^{(ph)}$ correspond to the particle-hole response, while those containing $R^{(pp)}$ and $R^{(hh)}$ are the analogous correlated propagators of two particles and two holes, respectively.
}
\label{SE2irrc}%
\end{figure}

\begin{figure}[t]
\begin{center}
\includegraphics[width = 0.6\textwidth]{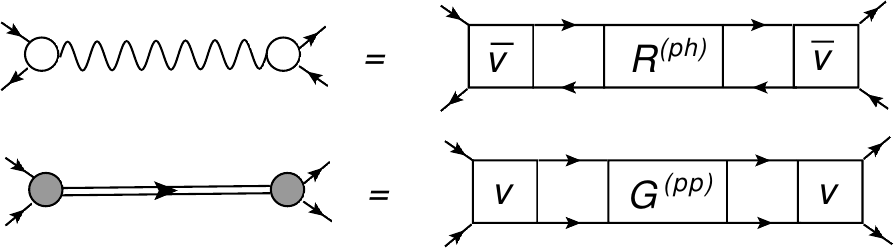}
\end{center}
\caption{The exact mapping of the phonon vertices (empty and filled circles) and propagators (wavy lines and double lines) onto the bare interaction (squares, antisymmetrized $\bar v$ and plain $v$) and two-fermion correlation functions (rectangular blocks $R^{(ph)}$ and $G^{(pp)}$) in a diagrammatic form. Lines with arrows stand for fermionic particles (right arrows) and holes (left arrows). Top: normal (particle-hole) phonon, bottom: pairing (particle-particle) phonon,  as introduced in Eq.~(\ref{vert_pp_ph}). The figure is adopted from Ref. \cite{LitvinovaSchuck2019}.}
\label{PVCmap}%
\end{figure}

(ii) Another possibility is to retain correlations in one of the two-body propagators in the factorization of the dynamical kernel. In this case, the irreducible correlated $2p2h$ propagator $G$ reads
\bea
&& G^{(c)irr}(543'1',5'4'31) \nonumber \\
&=& \langle T({\psi^{\dagger}}_1{\psi^{\dagger}}_3)(t)({\psi}_{3'}{\psi}_{1'})(t')\rangle \langle T({\psi}_5{\psi}_4)(t)({\psi^{\dagger}}_{4'}{\psi^{\dagger}}_{5'})(t')\rangle_0 \nonumber \\
&&+ \langle T({\psi^{\dagger}}_1{\psi^{\dagger}}_3)(t)({\psi}_{3'}{\psi}_{1'})(t')\rangle_0 \langle T({\psi}_5{\psi}_4)(t)({\psi^{\dagger}}_{4'}{\psi^{\dagger}}_{5'})(t')\rangle  \nonumber \\
&&+ \langle T({\psi^{\dagger}}_1{\psi}_5)(t)({\psi^{\dagger}}_{5'}{\psi}_{1'})(t')\rangle \langle T({\psi^{\dagger}}_3{\psi}_4)(t)({\psi^{\dagger}}_{4'}{\psi}_{3'})(t')\rangle_0  \nonumber\\
&&+ \langle T({\psi^{\dagger}}_1{\psi}_5)(t)({\psi^{\dagger}}_{5'}{\psi}_{1'})(t')\rangle_0 \langle T({\psi^{\dagger}}_3{\psi}_4)(t)({\psi^{\dagger}}_{4'}{\psi}_{3'})(t')\rangle   
- {AS},
\label{Gcirr}
\eea
where the upper index `(c)' indicates the presence of one two-fermion correlation function in each term of the expansion and `$AS$' stands for all the antisymmetric terms corresponding to the terms shown explicitly. This approximation to the dynamical kernel is illustrated diagrammatically in Fig.~\ref{SE2irrc} for the term $T^{(r;11)irr}_{12,1'2'}(t-t')$ of Eq.~(\ref{Tr11}).
It can be mapped to the class of approaches, which model the dynamical kernel in terms of the particle-vibration coupling. This mapping is exact, if the pairing $\gamma^{\mu{\pm}}$ and normal $g^{\mu{\pm}}$ phonon vertices as well as the respective propagators $\Delta_{\mu}^{(\pm)}$ and $D_{\mu}^{(\pm)}$ are defined as
\bea
\gamma^{\mu(+)}_{12} &=& \sum\limits_{34} v_{1234}\alpha_{34}^{\mu},\qquad\qquad\qquad \gamma_{12}^{\mu(-)} = \sum\limits_{34}\beta_{34}^{\mu}v_{3412},\nonumber \\
\Delta^{(\sigma)}_{\mu}(\omega) &=& \frac{\sigma}{\omega - \sigma(\omega_{\mu}^{(\sigma\sigma)} - i\delta)},\nonumber \\
g^{\nu(\sigma)}_{13} &=& \delta_{\sigma,+1}g^{\nu}_{13} + \delta_{\sigma,-1}g^{\nu\ast}_{31},\qquad\quad
g^{\nu}_{13} = \sum\limits_{24}{\bar v}_{1234}\rho^{\nu}_{42},\nonumber \\
D_{\nu}^{(\sigma)}(\omega) &=& \frac{\sigma}{\omega - \sigma(\omega_{\nu} - i\delta)}, \label{vert_pp_ph}
\eea 
with $\sigma = \pm 1$, via the normal and anomalous transition densities,
\be
\rho^{\nu}_{12} = \langle 0|\psi^{\dagger}_2\psi_1|\nu \rangle,\qquad
\alpha^{\mu}_{12} = \langle 0|\psi_2\psi_1|\mu\rangle,\qquad
\beta^{\mu}_{12} = \langle 0|\psi^{\dagger}_2\psi^{\dagger}_1|\mu\rangle.
\label{trden2}
\ee
This mapping is shown diagrammatically in Fig.~\ref{PVCmap}, and it expresses the mechanism of emergence of the collective degrees of freedom associated with phonons. One can see, for instance, that, if this mapping is applied to the first and the third terms on the right hand side of Fig.~\ref{SE2irrc}, together with an analogous mapping of the remaining components of $T^{(r)irr}$, the dynamical kernel takes the form of the conventional NFT-PVC approach \cite{BertschBortignonBroglia1983, KamerdzhievTertychnyiTselyaev1997, MahauxBortignonBrogliaEtAl1985,Tselyaev1989, LitvinovaRingTselyaev2008, LitvinovaWibowo2018, Tselyaev2016, NiuNiuColoEtAl2015, LitvinovaRingTselyaev2007, EgorovaLitvinova2016, RobinLitvinova2018}. 
The remaining terms, such as the second and fourth terms on the right hand side of Fig.~\ref{SE2irrc}, are not explicitly associated with the phonon exchange and represent a different type of correlations.

\begin{figure}[t]
\begin{center}
\includegraphics[width = \textwidth]{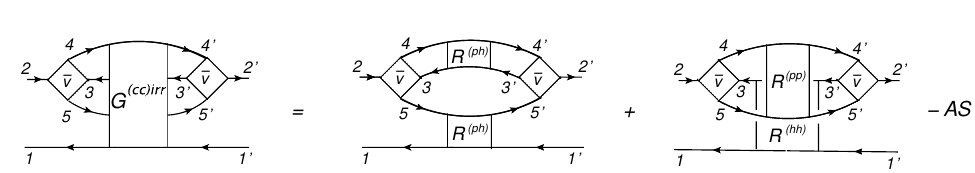}
\end{center}
\caption{Same as in Fig.~\ref{SE2irrc}, but for the doubly correlated approximation (iii).} 
\label{SE2irrcc}%
\end{figure}

(iii) More of the essential dynamics can be included if both of the pairwise propagators in the factorization of the dynamical kernel are exact and contain all correlations. The irreducible intermediate $2p2h$ propagator in this case reads
\bea
&&G^{(cc)irr}(543'1',5'4'31)  \nonumber \\
&=& \langle T({\psi^{\dagger}}_1{\psi^{\dagger}}_3)(t)({\psi}_{3'}{\psi}_{1'})(t')\rangle \langle T({\psi}_5{\psi}_4)(t)({\psi^{\dagger}}_{4'}{\psi^{\dagger}}_{5'})(t')\rangle  \nonumber \\
&&+ \langle T({\psi^{\dagger}}_1{\psi}_5)(t)({\psi^{\dagger}}_{5'}{\psi}_{1'})(t')\rangle \langle T({\psi^{\dagger}}_3{\psi}_4)(t)({\psi^{\dagger}}_{4'}{\psi}_{3'})(t')\rangle
- {AS},
\label{Gccirr}
\eea
and its diagrammatic image is given in Fig.~\ref{SE2irrcc}. Remarkably, this type of dynamical kernel absorbs the maximal amount of the correlations via the two-fermion correlation functions and it takes the simplest form among the three dynamical kernels defined by Eqs.~(\ref{G0irr}), (\ref{Gcirr}), and (\ref{Gccirr}).
It has the minimal amount of terms and eliminates the single-particle propagators. By the mapping shown in Fig.~\ref{PVCmap}, this kernel can be related with the QPM or multiphonon models  \cite{Soloviev1992,Ponomarev1999b,LoIudice2012,Lenske:2019ubp}
and with the extensions of the PVC approaches \cite{LitvinovaSchuck2019,LitvinovaRingTselyaev2010,Shen2020}.

It should be noted that most of the implementations of the dynamical kernels (i)-(iii) listed above used the effective interactions, instead of the bare interaction $\bar v$ and, simultaneously, the same effective interactions to approximate the entire static kernel. This means that in such approaches the many-body correlations are distributed differently between the static and dynamical kernels. Such a deficiency is, then, compensated by a subtraction procedure proposed in Ref.~\cite{Tselyaev2013}. On the other hand, the class of beyond-(Q)RPA approaches based on the bare interaction \cite{Knapp:2014xja,PapakonstantinouRoth2009,Bacca:2013dma,Raimondi:2018mtv,Bacca2014}
does not demonstrate a consistent performance of the quality comparable to that of the approaches employing the effective interactions. The listed \textit{ab-initio} 
implementations have a common feature that the dynamical kernels are not related to the static kernels as it is required by the exact EOM. Instead, various pre-processing methods, such as the in-medium similarity renormalization group \cite{Hergert2016} or Br\"uckner's G-matrix \cite{Dickhoff2004}, are applied to the bare interaction, with a subsequent application of one of the standard many-body methods.  


Another difficulty may arise from the use of the reference mean fields in the numerical implementation. Such auxiliary mean fields have to be subtracted from the interacting part of the Hamiltonian and, thus, go through all the commutators in both the static and dynamical kernels. This is done, in particular, for the static one-fermion kernel in Ref. \cite{Dickhoff2004}. However, new terms with few-fermion propagators appear in the dynamical kernels, that produces additional non-trivial non-linearities in the resulting EOMs. These terms and their roles in the emergent properties of the dynamical kernels have to be carefully analyzed, both analytically and quantitatively. 

\section{\textit{Nuclear spectral calculations}}

The response theory and the associated diagonalization approach are widely applied to the description of nuclear excited states. The response function (\ref{phresp}) is particularly convenient as it is directly related to the excitation energies and transition probabilities, which is obvious from its spectral expansion (\ref{respspec}). In experiments, the transition probabilities and their derivatives with respect to the energy variable are commonly extracted from the measured cross sections. The strength function as a result of the response to a given external field associated with the operator $F$ can be defined as
\be
S(\omega) = \sum\limits_{\nu>0} \Bigl[ |\langle \nu|F^{\dagger}|0\rangle |^2\delta(\omega-\omega_{\nu}) - |\langle \nu|F|0\rangle |^2\delta(\omega+\omega_{\nu})
\Bigr],
\label{SF}
\ee
where the summation over $\nu$ runs through all the excited states $|\nu\rangle$. The matrix element of the transition between the ground and excited states, in the case of one-body external field operator, reads
\be
\langle \nu|F^{\dagger}|0\rangle = \sum\limits_{12}\langle \nu|F_{12}^{\ast}\psi^{\dagger}_2\psi_1|0\rangle = \sum\limits_{12}F_{12}^{\ast}\rho_{21}^{\nu\ast}.
\ee
In the numerical implementations, the delta-functions in Eq.~(\ref{SF}) are approximated by the Lorentz distribution,
\be
\delta(\omega-\omega_{\nu}) = \frac{1}{\pi}\lim\limits_{\Delta \to 0}\frac{\Delta}{(\omega - \omega_{\nu})^2 + \Delta^2},
\ee
so that
\bea
S(\omega) &=& \frac{1}{\pi}\lim\limits_{\Delta \to 0}\sum\limits_{\nu} \Bigl[ |\langle \nu|F^{\dagger}|0\rangle |^2\frac{\Delta}{(\omega - \omega_{\nu})^2 + \Delta^2}
- |\langle \nu|F|0\rangle |^2\frac{\Delta}{(\omega + \omega_{\nu})^2 + \Delta^2}
\Bigr] \nonumber\\
&=& -\frac{1}{\pi}\lim\limits_{\Delta \to 0} {\Im}\sum\limits_{\nu} \Bigl[\frac{ |\langle \nu|F^{\dagger}|0\rangle |^2}{\omega - \omega_{\nu} + i\Delta}
- \frac{|\langle \nu|F|0\rangle |^2}{\omega + \omega_{\nu} + i\Delta}
\Bigr] \nonumber\\
&=& -\frac{1}{\pi}\lim\limits_{\Delta \to 0} {\Im} \Pi(\omega).
\label{SFDelta} 
\eea
The quantity $\Pi(\omega)$ is the polarizability of the many-body system,
\be
\Pi(\omega) = \sum\limits_{\nu} \Bigl[ \frac{|\langle \nu|F^{\dagger}|0\rangle |^2}{\omega - \omega_{\nu} + i\Delta}
- \frac{|\langle \nu|F|0\rangle |^2}{\omega + \omega_{\nu} + i\Delta}
\Bigr] 
=  \sum\limits_{\nu} \Bigl[ \frac{B_{\nu}}{\omega - \omega_{\nu} + i\Delta}
- \frac{{\bar B}_{\nu}}{\omega + \omega_{\nu} + i\Delta}
\Bigr]
\label{Polar}
\ee
with the transition probabilities defined as
\be
B_{\nu} = |\langle \nu|F^{\dagger}|0\rangle |^2,
\qquad\qquad
{\bar B}_{\nu} = |\langle \nu|F|0\rangle |^2.
\label{Prob}
\ee
Thus, the strength function associated with the given external field operator $F$ can be computed with the aid of the response function (\ref{phresp}):
\be
S_F(\omega) = -\frac{1}{\pi}\lim_{\Delta\to 0}\Im\sum\limits_{121'2'}F_{12}R_{12,1'2'}(\omega+i\Delta)F^{\ast}_{1'2'}.
\label{SFF}
\ee
In principle, the strength function (\ref{SFF}) should reproduce all the states excited by the operator $F$, if the response function (\ref{phresp}) is computed exactly.
The latter is, however, a difficult task, as it follows from its EOM (\ref{BSE-Dyson}) and from the explicit expressions for the respective interaction kernels $K^{(0)}$ and  $K^{(r)}$.  
As discussed before, in practice, various approximations are applied to these kernels. The finite imaginary part $\Delta$ of the energy variable introduces a smooth envelope of the strength function, otherwise it would look as a series of infinitely narrow peaks of infinite height. Smoothing of such a distribution is, thus, useful for representation purposes and has no physical meaning, if the theory is exact. Since the latter is practically never the case, the smearing parameter may absorb the effects that are not taken into account explicitly in the dynamical kernel $K^{(r)}$, and, thus, mimic the missing fragmentation effects. As the experimental data usually have finite energy resolution, the resulting spectral peaks also have finite widths and heights. The common agreement is, thus, that, for a fair comparison between theory and experiment, the smearing parameter $\Delta$ used in the calculations should be comparable with the experimental energy resolution. As the physical observable is the transition probability, its value should not depend on the smearing parameter. Indeed, at the peaks of the strength function, the following relationship holds,
\be
B_{\nu} = \lim_{\Delta\to 0} \pi\Delta S(\omega_{\nu}),
\label{BS}
\ee 
which points out that the choice of the smearing parameter does not affect the transition probabilities, although Eq. (\ref{BS}) gets, obviously, less accurate with larger $\Delta$.

In the next subsections, the authors overview some examples of realistic nuclear response calculations, on both the RPA and beyond-RPA levels. The major focus is put on the recent theoretical achievements in the description of nuclear collective modes, while comprehensive reviews including early studies can be found in Refs.~\cite{Ishkhanov2021,PaarVretenarKhanEtAl2007, Roca-Maza2018,BertschBortignonBroglia1983,KamerdzhievTertychnyiTselyaev1997,Drozdz:1990zz,KamerdzhievSpethTertychny2004,Garg2018}.

\paragraph{\textit{Electromagnetic and isoscalar response}}

The electromagnetic response is the most studied type of nuclear response as it can be induced by the most accessible experimental probes with photons \cite{Ishkhanov2021,GR2001,SavranAumannZilges2013}. The corresponding excitation operators are classified by the transferred angular momentum $L$ and parity $\pi$.
The electric operators have natural parity, i.e., $\pi = (-1)^L$, and are defined as
\bea
F_{00} &=& e\sum\limits_{i = 1}^Zr_i^2,\qquad F_{1M} = \frac{eN}{A}\sum\limits_{i=1}^Z r_iY_{1M}({\hat{\bf r}}_i) - \frac{eZ}{A}\sum\limits_{i=1}^N r_iY_{1M}({\hat{\bf r}}_i),
\nonumber\\
F_{LM} &=& e\sum\limits_{i = 1}^Zr_i^{\ L}Y_{LM}({\hat{\bf r}}_i),\qquad L \geq 2,
\label{FEL}
\eea
where $e$ stands for the proton charge, $Y_{LM}({\hat{\bf r}})$ are the spherical harmonics, and $Z$ and $N$ are the numbers of protons and neutrons in a nucleus, respectively. The expression for $L = 1$ contains the ``kinematic'' charges to account for the center-of-mass motion. Otherwise, the electric excitation operators (\ref{FEL}) imply only the interaction of the projectiles with the charged protons and no interaction with the neutrons. The corresponding isoscalar operators with zero isospin transfer contain summations over all the nucleons and no electric charge, if they are not associated with the electric probes. 
The isoscalar dipole operator reads
\be
F^{(0)}_{1M} = \sum\limits_{i=1}^A (r^3_i - \eta r_i)Y_{1M}({\hat{\bf r}}_i),  
\label{opISE1}
\ee 
where $\eta = 5\langle r^2\rangle/3$, and the second term in the brackets eliminates the spurious translational mode \cite{Garg2018}. The superscript `(0)' indicates the isoscalar character of the operator, $\Delta T = 0$, in contrast to the operators (\ref{FEL}), which are often classified as isovector ones with $\Delta T = 1$.
The magnetic multipole operators are of the unnatural parity $\pi = (-1)^{L+1}$ and of a more complex nature. 
Magnetic resonances are associated with the spin transfer and, generally, do not exhibit pronounced collectivity \cite{Tselyaev2020}.
Therefore, the authors focus on the electric multipole transitions in this section.

The response of strongly-correlated systems to external perturbations manifests some generic features of the excitation spectra, which can be captured by a schematic model 
proposed by Brown and Bolsterli \cite{BrownBolsterli1959,RingSchuck1980}. In this model, which adopts a separable effective multipole-multipole interaction, 
the RPA excitation spectrum contains two highly collective states, the low-frequency and the high-frequency ones. These two states are formed by the coherent particle-hole contributions from the uncorrelated $ph$-excitations (with respect to the Hartree-Fock or the phenomenological mean-field vacuum), when the interaction is switched on. The remaining $ph$ states are mostly non-collective and lie between the two collective solutions. In the RPA calculations with more realistic interactions, the resulting spectrum depends on the nature of the residual interaction and on the quality of the numerical implementation. The general gross structure of the spectrum remains as in the Brown-Bolsterli model, but the main collective solutions undergo some fragmentation, the so-called Landau damping. Typically, the low-energy solutions are not very collective in the $L = 0$ and $L =1$ channels, but acquire collectivity at larger angular momentum transfer $L \geq 2$.
The high-energy ones are associated with collective oscillations, which involve all the nucleons. Taking into account the dynamical kernel, in any of the approximations discussed above, induces further fragmentation of the $ph$ states due to their coupling to more complex configurations. This effect is a consequence of the pole structure of the dynamical kernel. 
The fine details of the obtained spectra vary depending on the approximation to $K^{(r)}$. 

\begin{figure}
\begin{center}
\includegraphics[width=\textwidth]{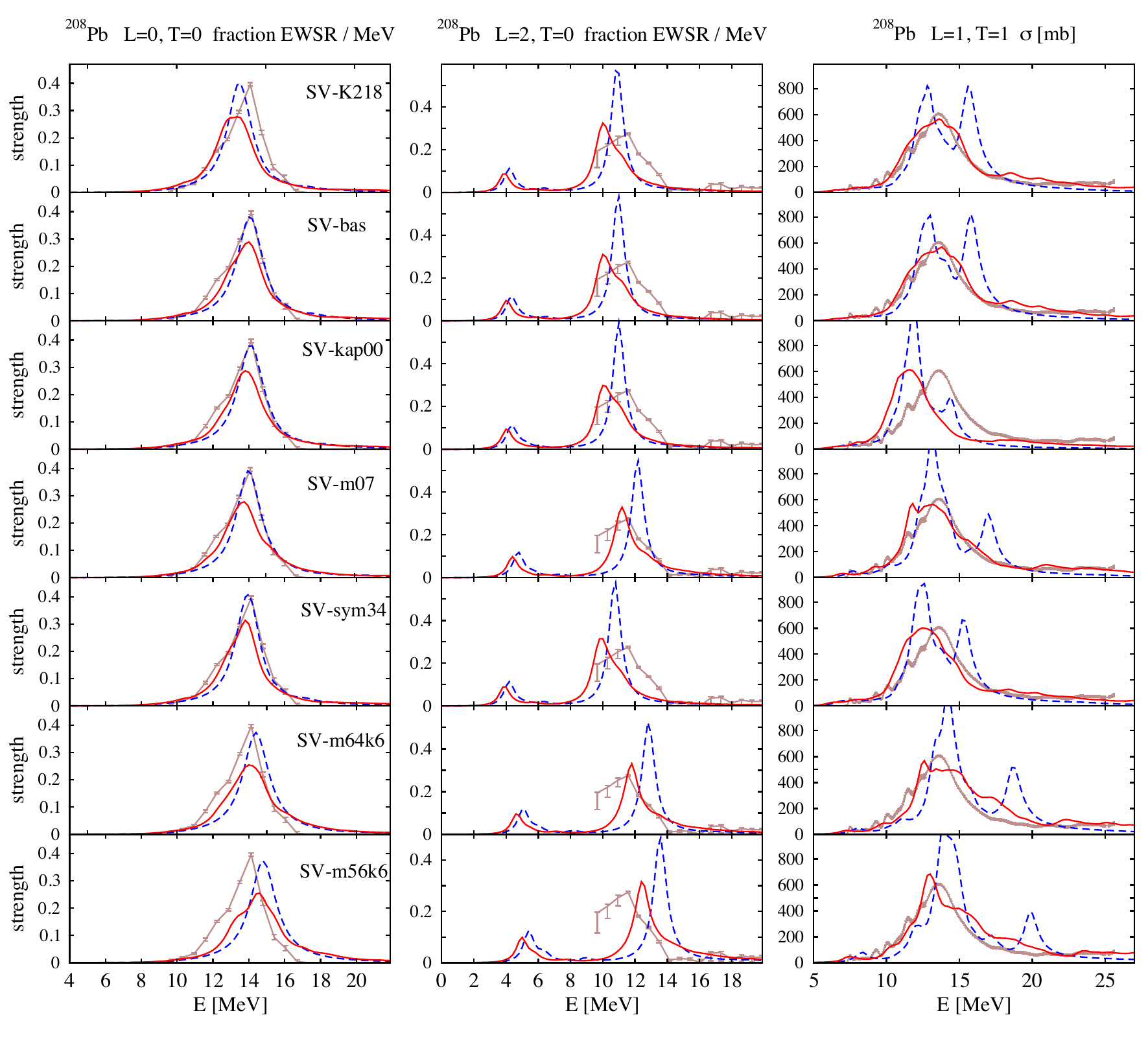}
\end{center}
\vspace{-0.3cm}
\caption{Isoscalar monopole (left), isoscalar
quadrupole (middle), and isovector dipole (right) strength distributions in $^{208}$Pb in the units of the energy-weighted sum rules per MeV for $L = 0$ and $L = 2$, and as a photoabsorption cross section [$\sigma(E)$ defined in Eq.~(\ref{cs})] for the dipole mode. The RPA results are shown by the blue dashed lines, and the beyond-RPA PVC extensions are given by the red solid lines. The experimental data for the GDR \cite{Belyaev1995} and for the GMR and GQR \cite{Youngblood2004} are shown by the brown lines with error bars. The figure is adopted from Ref.~\cite{Tselyaev2016}.}
\label{gr-208Pb}%
\end{figure}

Figure~\ref{gr-208Pb} illustrates  microscopic calculations of the isoscalar monopole, isoscalar
quadrupole, and isovector dipole responses in $^{208}$Pb of Ref.~\cite{Tselyaev2016}. The strength distributions were obtained with various Skyrme interactions in both the RPA and beyond-RPA model that includes the particle-vibration coupling in the time blocking approximation (TBA), which results in the PVC dynamical kernel of the NFT type (ii). The calculations are done with a relatively large smearing parameter of the order of $1$~MeV. The case of the quadrupole response is a clear illustration of the Brown-Bolsterli picture, while in the dipole and monopole channels the low-energy peak is not distinguishable. It is possibly too weak in the electromagnetic dipole channel, while in the monopole case it can be suppressed by the monopole selection rule. The giant resonances at high frequencies are, however, well pronounced showing up as broad peaks dominating the spectra. The fragmentation due to PVC shows up as a broadening of the giant resonance also in all the three channels, however, the effect is weaker for the monopole response. The latter occurs due to the partial cancellation between the self-energy $K^{(r;11)}, K^{(r;22)}$ and phonon-exchange $K^{(r;12)}, K^{(r;21)}$ terms, that is typical for the PVC kernels in the $L = 0$ channel \cite{BBB1998,LitvinovaRingTselyaev2007}. The use of a large smearing parameter in the calculations for the dipole response to reproduce the data, as compared with the experimental energy resolution, indicates the deficiency of fragmentation originating from the PVC mechanism. This can be attributed to the underestimated phonon collectivity in the Skyrme-RPA calculations employed for obtaining the phonon characteristics and/or to the deficiency of the PVC model space. While the choice of the smearing parameter looks more adequate with respect to the experimental resolution in the $L = 0$ and $L = 2$ cases, the RPA-PVC calculations underestimate the peak height of the GMR and, in some cases, the centroid of the GQR. Thus, the currently available Skyrme-RPA-PVC results for $^{208}$Pb call for further refinement of the Skyrme interactions and/or of the employed many-body calculation schemes.

\begin{figure}[t]
\begin{center}
\includegraphics[width=0.95\textwidth]{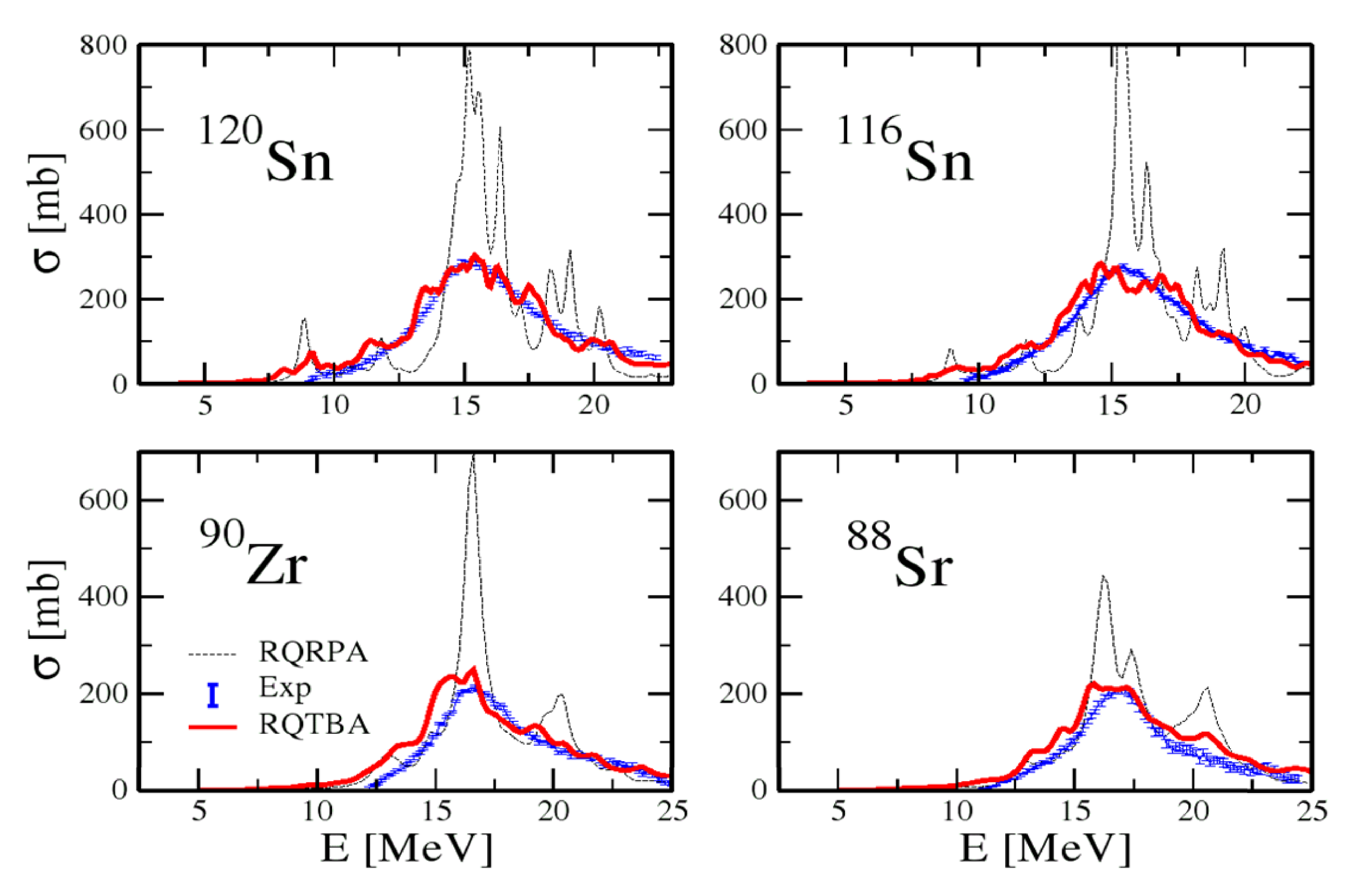}
\end{center}
\vspace{-0.3cm}
\caption{Total dipole photoabsorption cross section in stable medium-mass nuclei. The figure is adopted from Refs. \cite{50BCS,Meng2016}.}
\label{gdr}%
\end{figure}

Figure~\ref{gdr} shows the cross sections of the total dipole photoabsorption in four medium-mass spherical nuclei obtained within the relativistic QRPA \cite{PaarRingNiksicEtAl2003} (RQRPA, black dashed curves) and the relativistic quasiparticle TBA \cite{LitvinovaRingTselyaev2008} (RQTBA, red solid curves), compared with the neutron data (blue error bars) from Ref.~\cite{nndc}. This cross section is defined as 
\be
\sigma_{E1}(E)={\frac{{16\pi^{3}e^{2}}}{{9\hbar c}}}E~S_{E1}(E),
\label{cs}
\ee
i.e., with the additional energy factor in front of the strength distribution, which slightly emphasizes the high-energy part of the response.
These calculations also employ the PVC dynamical kernel in its NFT form, which has been generalized to the superfluid phase, i.e., to the coupling between the superfluid quasiparticles and phonons \cite{LitvinovaRingTselyaev2008,LitvinovaTselyaev2007}. The in-medium interaction is of the effective meson-exchange origin and adjusted to bulk nuclear properties in the framework of the covariant DFT \cite{VretenarAfanasjevLalazissisEtAl2005, Meng2006, Meng2016} with the NL3 parametrization \cite{NL3}. In the fully self-consistent calculation scheme, the RQRPA generally produces the dipole strength, which is  mostly concentrated in a narrow energy region. The localization of the centroid is reproduced fairly well, as compared with the data. 

The total transition probability is another characteristic of the GDR, which is typically reproduced well in the (Q)RPA approaches. The most robust related quantity is the energy-weighted sum rule (EWSR),
\be
S_{E1} = \sum\limits_{\nu} E_{\nu}B_{\nu} = \frac{9\hbar^2e^2}{8m_p}\frac{NZ}{A},
\label{sr}
\ee
which is proportional to the cross section integrated over the energy variable. The right hand side of Eq.~(\ref{sr}) is calculated by transforming the sum into a double commutator of the dipole excitation operator and the system Hamiltonian, under the assumption that the interaction between nucleons has no momentum dependence. In this case, the potential energy part commutes with the excitation operator and, thus, does not contribute to the sum rule. The relation~(\ref{sr}) is, therefore, valid for any Hamiltonian without momentum dependence in the two-body sector and known as Thomas-Reiche-Kuhn sum rule. Modern energy density functionals (EDFs), such as the Skyrme, Gogny, and relativistic ones yield the effective interactions, which depend on the nucleonic momenta, so that a $10$--$20\%$ or even larger enhancement of the dipole EWSR can be obtained in the (Q)RPA calculations \cite{Trippa2008} as well as in experiments, where the measurements span sufficiently broad energy intervals \cite{nndc}. 

Adding the dynamical kernels, which satisfy the consistency conditions between the self-energy $K^{(r;11)}, K^{(r;22)}$ and the exchange $K^{(r;12)}, K^{(r;21)}$ terms, 
should not violate the EWSR \cite{Tselyaev2007}. In particular, the (quasi)particle-vibration coupling (QPVC) kernels of the NFT form (ii) satisfy this condition, if the numerical implementation is performed properly. Thus, the EWSR conservation serves as a very good test for such implementations. Accordingly, the energy centroid remains intact. The subtraction procedure \cite{Tselyaev2013}, which is applied to eliminate the double counting of the QPVC effects in EDFs, induces a slight violation of the EWSR, because it modifies the static part of the kernel and pushes the centroid slightly upward, so that the resulting position of the major peak is back to its (Q)RPA position. Otherwise, the dynamical kernel alone shifts the major peak to lower energy. This is a desirable feature in the \textit{ab-initio} implementations, such as the second RPA of Ref.~\cite{PapakonstantinouRoth2009}. However, if an effective interaction is employed for the dynamical kernel, the major peak is already well positioned in (Q)RPA, so that its downward shift by the dynamical kernel is well compensated by the subtraction. This procedure is quite simple and consists of the replacement,
\be
{\tilde K}^{(0)} + {\tilde K}^{(r)}(\omega) \to {\tilde K}^{(0)}  + \delta {\tilde K}^{(r)}(\omega) = {\tilde K^{(0)}}  + {\tilde K}^{(r)}(\omega) - {\tilde K}^{(r)}(0), 
\label{subtraction}
\ee
i.e., the dynamical kernel in the static approximation $\omega = 0$ is subtracted from the dynamical kernel itself. The energy-independent combination ${\tilde K}^{(0)} - {\tilde K}^{(r)}(0)$, thus, stands for the effective interaction freed from the long-range effects taken into account by  ${\tilde K}^{(r)}(\omega)$. The `$\ \widetilde{ }$\ ' sign in Eq.~(\ref{subtraction}) marks the kernels, where the effective interaction is employed, that is, the entire static kernels ${\tilde K}^{(0)}$ and the interaction matrix elements in the topologically equivalent dynamical kernels ${\tilde K}^{(r)}(\omega)$ computed in various approximations.

One can further see from Figure~\ref{gdr}, that the coupling between the superfluid quasiparticles and phonons included within the RQTBA provides a sizable fragmentation of the GDR. Due to the inclusion of a large number of the phonon modes, the final strength distribution acquires nearly a Lorentzian shape, though relatively small values of the smearing parameter, $\Delta = 200$~keV for the Sn isotopes and $\Delta = 400$~keV for Sr and Zr, were used in both the RQRPA and RQTBA calculations. The choice of these parameters was based on the estimate of the continuum contribution, which was not included explicitly. 
 
In principle, the particle escaping to the continuum plays a role in the formation of the width of the high-frequency resonances above the particle emission threshold. The latter is the minimal energy, at which the nucleon emission is possible, often called nucleon binding, or separation, energy, and its typical value is $\sim 7$--$10$~MeV for stable medium-mass and heavy nuclei.  Loosely-bound exotic nuclei with strong dominance of one type of nucleons (protons or neutrons) are characterized by lower separation energies for the excess nucleons. For example, in neutron-rich nuclei, neutrons are loosely bound and have lower separation energy than protons, and vice versa. The effect of the single-particle continuum in the (Q)RPA and beyond-(Q)RPA calculations can be taken into account within the method first proposed in Ref.~\cite{ShlomoBertsch1975} for RPA, later extended to QRPA \cite{Tselyaev2016,Kamerdzhiev1998,HaginoSagawa2001,Matsuo2002,KhanSandulescuGrassoEtAl2002,Daoutidis2009} and QRPA+QPVC \cite{LitvinovaTselyaev2007}. The complete inclusion of the single-particle continuum in these methods is achieved by employing the coordinate-space representation for the (Q)RPA propagator and the final EOM, while the QPVC part of the propagator in Ref.~\cite{LitvinovaTselyaev2007} is transformed to the coordinate space via the single-particle wave functions. In Ref.~\cite{Tselyaev2016}, a modification of this method was proposed for the numerical solution of the response EOM in the discrete basis of the single-particle states with the box boundary condition. Both the original and modified methods are based on constructing the mean-field propagator from the regular and irregular single-particle wave functions as the mean-field solutions with the Coulomb asymptotics.  

The single-particle continuum included in the calculations presented in Fig.~\ref{gr-208Pb}  does not play a very important role in the description of medium-mass and heavy nuclei, producing a typical continuum width of the order of $100$~keV for each single peak in the spectrum above the particle threshold, although the role of continuum increases dramatically in light nuclei, especially the loosely bound ones. Clear examples are given in Ref.~\cite{Tselyaev2016}. The inclusion of multiparticle continuum in (Q)RPA and its extensions was not addressed in the nuclear physics literature until now, though effects of two-nucleon evaporation should become sensible already at the GDR centroid energy and further escape of more nucleons can affect the GDR's high-energy shoulder. 
Further details of the calculations presented in Figs.~\ref{gr-208Pb} and \ref{gdr} can be found in Refs.~\cite{LitvinovaRingTselyaev2008,Tselyaev2016}, respectively.

%
\begin{figure}[t]
\begin{center}
\includegraphics[width=0.9\textwidth]{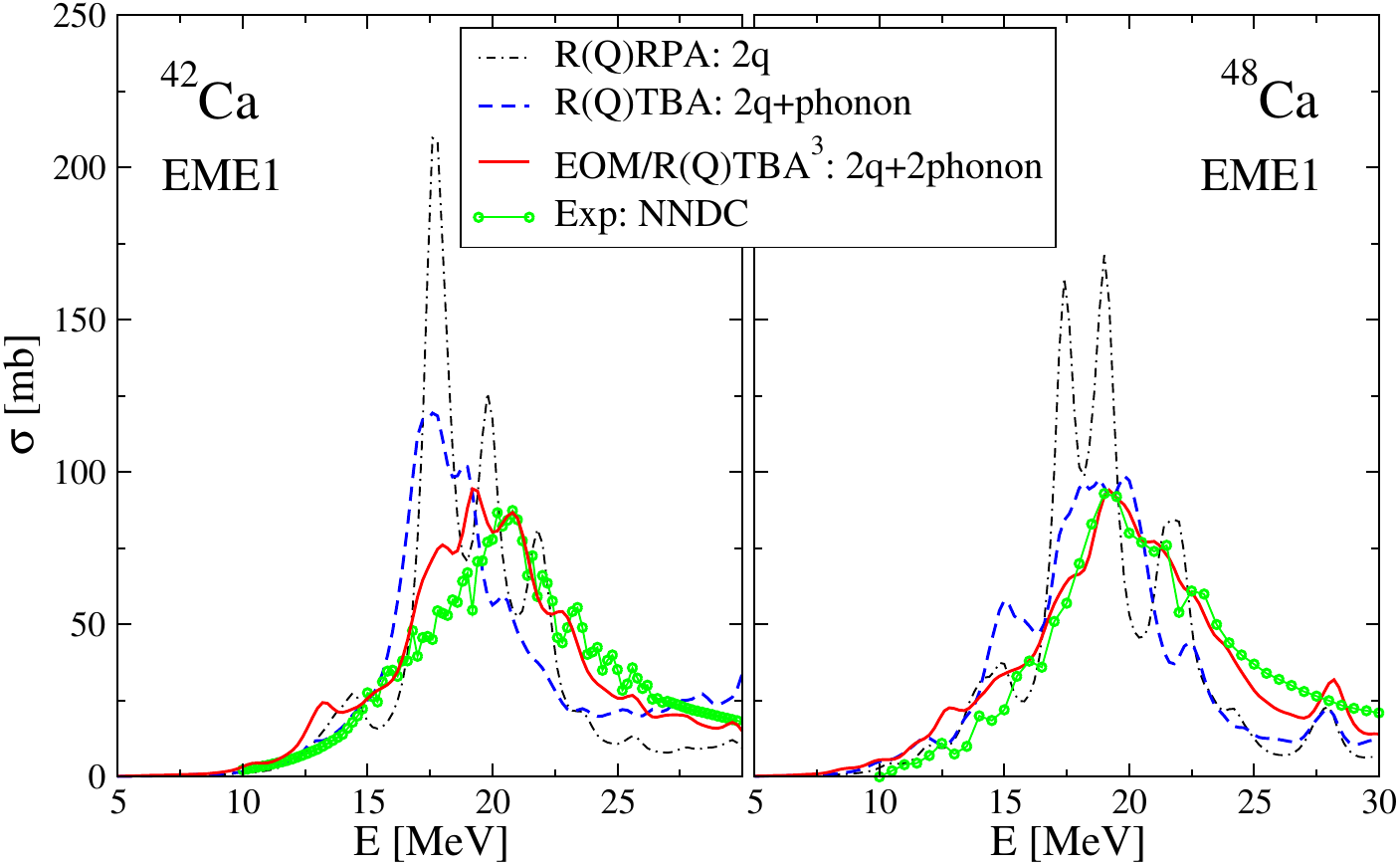}
\end{center}
\caption{Giant dipole resonance in $^{42,48}$Ca calculated within the R(Q)RPA, R(Q)TBA, and EOM/R(Q)TBA$^3$ approaches \cite{LitvinovaSchuck2019}, in comparison with the experimental data \cite{Erokhova2003,nndc}. The figure is adopted from Ref. \cite{LitvinovaSchuck2019}.}
\label{42-48-ca_gdr}%
\end{figure}
%

The described models with various types of dynamical kernels, although quite successful, still have not reached the spectroscopic accuracy of even hundreds of keV in the description of excitation spectra and other properties of medium-mass and heavy nuclei, which can be associated with these spectra. Despite the convincing progress on both the beyond-(Q)RPA methods and the EDFs, it remains unclear to what degree the lack of accuracy should be attributed to the imperfections of the EDFs, truncations in the beyond-(Q)RPA calculation schemes, unavoidable with the present computational capabilities, or principal limitations of these many-body methods. Up until now,  the best-quality nuclear response calculations beyond (Q)RPA include up to the (correlated) $2p2h$ \cite{LitvinovaRingTselyaev2008,LitvinovaRingTselyaev2010,NiuNiuColoEtAl2015,Gambacurta2015,RobinLitvinova2016,Robin2019}, in rare cases $3p3h$ \cite{LoIudice2012,LitvinovaSchuck2019,Lenske:2019ubp,
Ponomarev1999,Savran2011}, configuration complexity with the current computational capabilities. Direct comparison between the $2p2h$ and $3p3h$ calculations within the same implementation schemes indicates that the latter higher-rank configurations (i) improve the results noticeably and (ii) the effect of the inclusion of $3p3h$ configurations, in addition to $2p2h$ ones, is weaker than the effect of the inclusion of $2p2h$ configurations beyond (Q)RPA. The former points to the importance of the $3p3h$ configurations and the latter means that the theory exhibits saturation with respect to the configuration complexity. 

An example is given in Fig.~\ref{42-48-ca_gdr}, where $3p3h$ configurations were included in the "two quasiparticles coupled to two phonons" ($2q\otimes 2phonon$) scheme for the electromagnetic dipole response of $^{42,48}$Ca. This was achieved by implementing the dynamical QPVC kernel of type (iii) in an iterative cycle. Namely, after computing and selecting the most relevant RQRPA phonon modes (without dynamical kernels), the dynamical QPVC kernel (ii) was constructed, and the RQTBA response was calculated for the most relevant $J^{\pi}$ ($J\leq 6$) channels of natural parity. After that, the obtained response functions were recycled in the dynamical  kernel (iii) of the EOM for the dipole response. This scheme was originally proposed in Ref. \cite{Litvinova2015} within the quasiparticle time blocking approximation, and later re-derived starting from the bare Hamiltonian and implemented numerically in Ref. \cite{LitvinovaSchuck2019}. The approach was named EOM/RQTBA$^3$ due to its construction.
The total photoabsorption cross section obtained within EOM/RQTBA$^3$ (red solid curves) is plotted in Fig.~\ref{42-48-ca_gdr} together with the results of RQRPA (black dot-dashed curves), RQTBA (blue dashed curves) and experimental data (green curves and circles) of Ref. \cite{nndc}. 

The GDR in calcium isotopes was investigated within the RQTBA framework in Ref. \cite{EgorovaLitvinova2016} with the focus on the role of the $2q\otimes phonon$ configurations in the width of the GDR. It was found that these configurations result in the formation of the spreading width and improve significantly the agreement to data as compared to RQRPA. Nevertheless, although the authors used a large model space of the  $2q\otimes phonon$ configurations with the RQRPA phonons, the total width of the GDR was still underestimated.  In addition, on the high-energy shoulder of the GDR the cross sections were systematically underestimated. A similar situation was reported in Ref. \cite{Tselyaev2016}, for QTBA calculations with various Skyrme forces.  These observations pointed out that further refinement of the dynamical kernels may be necessary. Now with the EOM/RQTBA$^3$, taking into account more complex $2q\otimes 2phonon$ configuration, we can see that  these problems can be potentially resolved. Indeed, from Fig. \ref{42-48-ca_gdr} one can see that the new higher-rank configurations in EOM/RQTBA$^3$ cause additional fragmentation of the GDR and, thus, intensify the spreading of the strength to both higher and lower energies. Technically, this is the consequence of the appearance of  the new poles in the resulting response function. These new poles rearrange the energy balance of the strength distribution in both the low-energy and the higher-energy sectors, however, without violating the dipole EWSR \cite{LitvinovaSchuck2019}. 
\begin{figure}
\begin{center}
\vspace{-0.2cm}
\includegraphics[scale=0.32]{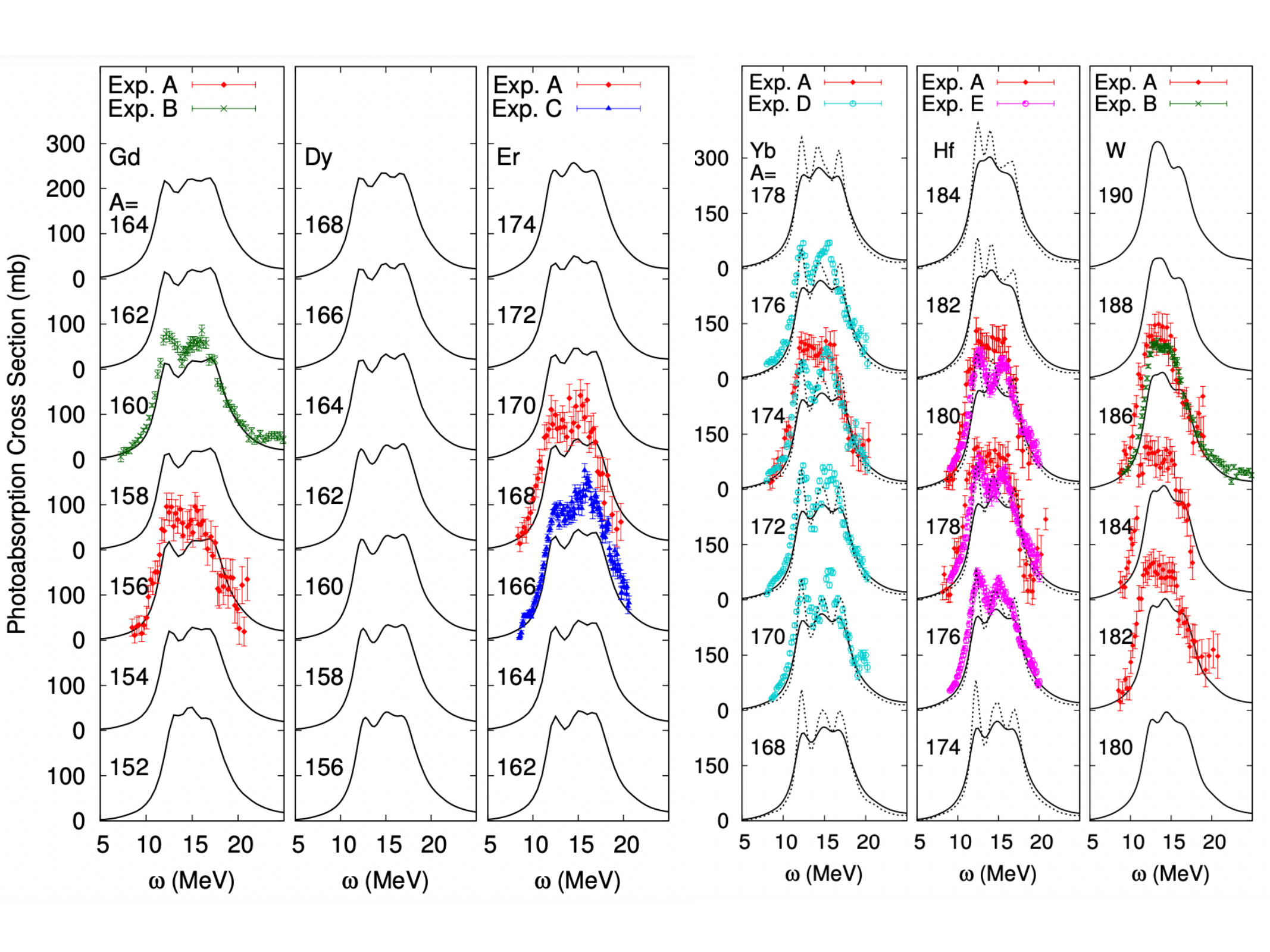}
\end{center}
\vspace{-0.3cm}
\caption{Dipole photoabsorption cross sections for Gd, Dy, and Er (left), and for Yb, Hf, and W (right) isotopes as a function of photon energy, obtained in the FAM-QRPA calculation with the Skyrme interaction and the smearing parameter $\Delta = 1.0$~MeV. Calculations with $\Delta = 0.5$~MeV are given for comparison (dotted lines). The figure is adopted from Ref. 
\cite{OishiKortelainenHinohara2016}.}
\vspace{-0.5cm}
\label{FAM-QRPA}%
\end{figure}

Response of non-spherical nuclei to external probes, in general, is more difficult to calculate  microscopically. Already on the QRPA level, the $2q$ model space expands dramatically, as compared to the spherical case. The reason is the lifted degeneracy of $j$-orbitals, because the total angular momentum is not a good quantum number in non-spherical geometries.  Therefore, QRPA calculations are numerically very expensive even in axially deformed nuclei \cite{PenaRing2008}. In particular, such calculations require numerical evaluation of the enormous amount of matrix elements of the nucleon-nucleon interaction, which makes deformed QRPA prohibitively difficult even in the DFT frameworks.
See also Refs.~\cite{Peru2008Phys.Rev.C77.044313, Toivanen2010Phys.Rev.C81.034312} for the studies along this direction.
A very elegant numerical solution was proposed in Ref.~\cite{Nakatsukasa2007}, where the finite-amplitude method (FAM), avoiding direct computation of the interaction matrix elements, was developed and employed for RPA calculations of the response of deformed nuclei. Later on, the FAM-RPA was generalized to superfluid nuclei as FAM-QRPA
 \cite{OishiKortelainenHinohara2016,Niksic2013,Kortelainen2015}. 

An example of FAM-QRPA calculations for the GDR in axially deformed nuclei is shown in Fig. \ref{FAM-QRPA}, in terms of the total dipole photoabsorption cross sections \cite{OishiKortelainenHinohara2016}. The calculations were performed with the Skyrme DFT. The experimental data on the GDR are provided by the authors of Ref.  \cite{OishiKortelainenHinohara2016} for the nuclei, where such data are available. In general, the double-hump shape of the GDR in axially-deformed nuclei is attributed to the deformation, while the ratio of the peak energies corresponds to the ratio of the major axes of the nuclear ground state ellipsoid.
The main observation from the Skyrme FAM-QRPA calculations shown in Fig. \ref{FAM-QRPA} is that the typical frequencies of the GDR are fairly well reproduced. The width and the plateau top of the distribution are well understood as the total $J = 1$ strength is a sum of $K = 0$ and $|K| = 1$ modes excited on ground states with  prolate deformations. 
The remaining discrepancies between the QRPA calculations and experimental data were attributed by the authors to the peculiarities of the Skyrme interaction and to the missing effects beyond QRPA. Indeed, since considerably larger smearing than the experimental energy resolution is needed in these FAM-QRPA calculations to reproduce the data, the effects beyond QRPA are necessary for further improvements. The work in this direction is under way \cite{Zhang2021}. 

\paragraph{\textit{Spin-isospin response}}

The nuclear spin-isopin response, also known as the charge-exchange excitations, corresponds to the transitions from the ground-state of the nucleus $(N,Z)$ to the final states in the neighboring nuclei $(N\mp1,Z\pm1)$ in the isospin lowering $T_-$ and raising $T_+$ channels, respectively.
These excitations can take place spontaneously, e.g., in the famous $\beta$ decays, or be induced by external fields, e.g., in the charge-exchange reactions, such as $(p, n)$ or $(^3{\rm He}, t)$.
Nuclear spin-isopin responses are categorized into different modes according to the nucleons with spin-up and spin-down oscillating either in phase, the non-spin-flip modes with $S = 0$, or out of phase, the spin-flip modes with $S = 1$. The important modes, which have attracted an extensive attention experimentally and theoretically, include the isobaric analog state with $S = 0$, $J^\pi = 0^{+}$, Gamow-Teller resonance with $S = 1$, $J^\pi = 1^{+}$, and spin-dipole resonance with $S = 1$, $J^\pi = 0^{-}, 1^{-}, 2^{-}$ \cite{Osterfeld1992, Ichimura2006, PaarVretenarKhanEtAl2007, Roca-Maza2018}.

The corresponding operators of these charge-exchange excitations read
\bea
F_{\rm IAS}^{\pm} &=& \sum_{i=1}^{A} \tau_\pm(i), \nonumber\\
F_{\rm GTR}^{\pm} &=& \sum_{i=1}^{A} [1 \otimes \overrightarrow{\sigma}(i)]_{J=1} \tau_\pm(i), \nonumber\\
F_{\rm SDR}^{\pm} &=& \sum_{i=1}^{A} [r_i Y_1(i) \otimes \overrightarrow{\sigma}(i)]_{J=(0,1,2)} \tau_\pm(i),
\eea
where $Y$ is the spherical harmonics, $\sigma$ and $\tau$ are the Pauli matrices of spin and isospin degrees of freedom, respectively.
The corresponding non-energy-weighted sum rules (NEWSR), $S^- - S^+ = \sum_\nu B^-_\nu - \sum_\nu B^+_\nu$, are
\bea
  S^-_{\rm IAS} - S^+_{\rm IAS} &=& N - Z, \nonumber\\
  S^-_{\rm GTR} - S^+_{\rm GTR} &=& 3(N - Z), \nonumber\\
  S^-_{\rm SDR} - S^+_{\rm SDR} &=& \frac{9}{4\pi}\Bigr[N\langle r^2\rangle_n - Z\langle r^2\rangle_p\Bigr],  
\eea
where the GTR one is the famous model-independent Ikeda sum rule, while the SDR one involves the root-mean-square radii of protons and neutrons and is considered to be an alternative way for measuring neutron skin thickness \cite{Krasznahorkay1999, Yako2006}.
For neutron-rich nuclei, the excitations in the $T_+$ channel are significantly suppressed by the Pauli principle, and thus $S^-$ alone approximately represents the NEWSR.

The GTR, which is the most studied nuclear spin-isospin response, is related to both the spin-orbit and isospin properties of nuclear systems. Although this relationship is not direct and clouded by complex many-body correlations, experimental data on the GTR can be used to constrain the respective terms in the effective interactions and EDFs.
For instance, one of the recently developed and widely used Skyrme effective interactions, SAMi \cite{Roca-Maza2012}, has acquired improved spin-isospin properties by achieving an accurate description of GTR peak energies. Meanwhile, in the relativistic framework, it is found that on the (Q)RPA level an accurate description of GTR peak energies can be achieved in a fully self-consistent way by taking the Fock terms of the meson-exchange interactions into account \cite{LiangVanGiaiMeng2008, Niu2013, Niu2017}.


Overall, RPA and QRPA with effective interactions \cite{Borzov:2003bb,Sarriguren2013,PaarNiksicVretenarEtAl2004a,LiangVanGiaiMeng2008, Niu2013, Niu2017} produce reasonable results for the major GTR peak. However, reproducing the detailed strength distribution is impossible within these approaches neglecting the dynamical correlations. Moreover, since the total strength is constrained by the model-independent Ikeda sum rule, it is exhausted within the relatively narrow energy interval, because of the model space limitations of (Q)RPA. This causes, to a large extent, the well-known quenching problem \cite{Osterfeld1992}. 
The overall situation is similar to that with electromagnetic excitations, and the GTR is considerably affected by the effects beyond RPA. While SRPA calculations for the GTR  have been reported already in 1990 \cite{Drozdz:1990zz}, calculations with the (Q)PVC kernels based on modern density functionals, both relativistic NL3 \cite{MarketinLitvinovaVretenarEtAl2012,
RobinLitvinova2016,RobinLitvinova2018,Robin2019} and non-relativistic Skyrme \cite{NiuNiuColoEtAl2015}, have become available more recently. Lately, SRPA calculations were also advanced to the Skyrme EDF framework \cite{Gambacurta2020}. Despite technical differences between the various implementations, all the extensions beyond (Q)RPA improve the description of the GTR considerably. In the cases of neutron-rich nuclei, where the low-energy part of the GTR spectrum is associated with spontaneous beta decay, the description of beta decay rates are improved by up to one or  two orders of magnitude, compared to those obtained in (Q)RPA \cite{RobinLitvinova2016,LitvinovaRobinWibowo2020}.

\begin{figure*}[ptb]
\begin{center}
\includegraphics[scale=0.45]{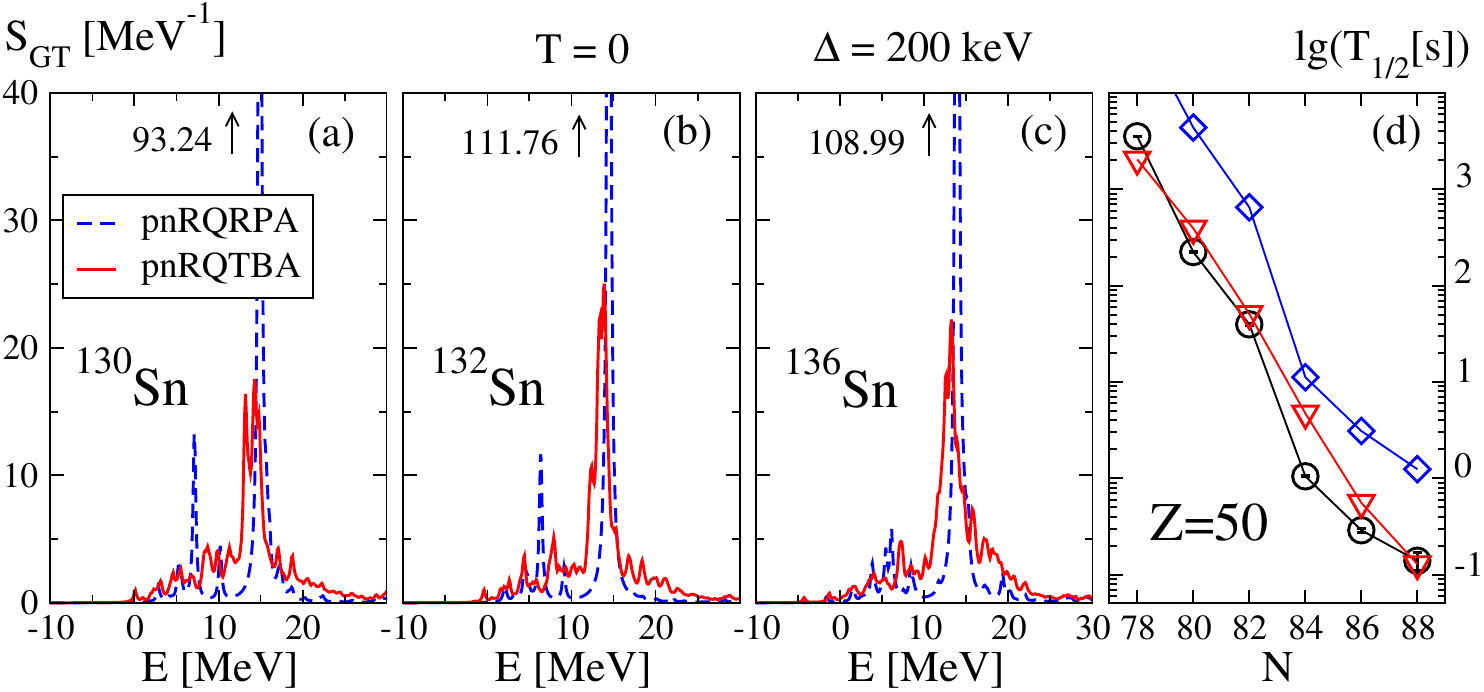}
\end{center}
\vspace{-0.2cm}
\caption{GT$_-$ strength distribution for $^{130,132,136}$Sn nuclei at zero temperature in the pnRQTBA, compared to the pnRQRPA (a-c). Beta decay half-lives in neutron-rich tin isotopes extracted from the pnRQRPA (diamonds) and pnRQTBA (triangles) strength distributions, compared to data (circles) \cite{nndc} (d). The figure is adopted from Ref. \cite{LitvinovaRobinWibowo2020}.}  
\label{GTR_Sn}
\end{figure*}

The role of QPVC effects is illustrated in Fig. \ref{GTR_Sn} for the response of the neutron-rich tin isotopes $^{130,132,136}$Sn to the GT$_-$ operator, obtained within the proton-neutron version of RQTBA (pnRQTBA) with the QPVC dynamical kernel, which was originally developed in Ref. \cite{RobinLitvinova2016}. These calculations are compared to the proton-neutron RQRPA (pnRQRPA) without the dynamical kernel, and the presented spectra are displayed on the energy scales relative to the parent nuclei. 
The most general observation from these calculations is that QPVC leads to a similar degree of fragmentation as in non-charge-exchange channels, which is somewhat higher in nuclei with larger isospin asymmetry. In turn,  this fragmentation redistributes the strength in the low-energy sector, in particular, in the $Q_{\beta}$ energy window.  This leads to faster beta decay in the pnRQTBA calculations, improving significantly the agreement with experimental data \cite{nndc}, as compared to pnRQRPA. The corresponding half-lives are shown in the right panel of Fig. \ref{GTR_Sn}. More examples, details and discussions are presented in Ref. \cite{RobinLitvinova2016}. Furthermore, the pnRTBA has been generalized recently to finite temperature in Ref. \cite{LitvinovaRobinWibowo2020}, which allowed applications to beta decay in stellar environments. 
\begin{figure}[t]
\begin{center}
\includegraphics[width=0.9\textwidth]{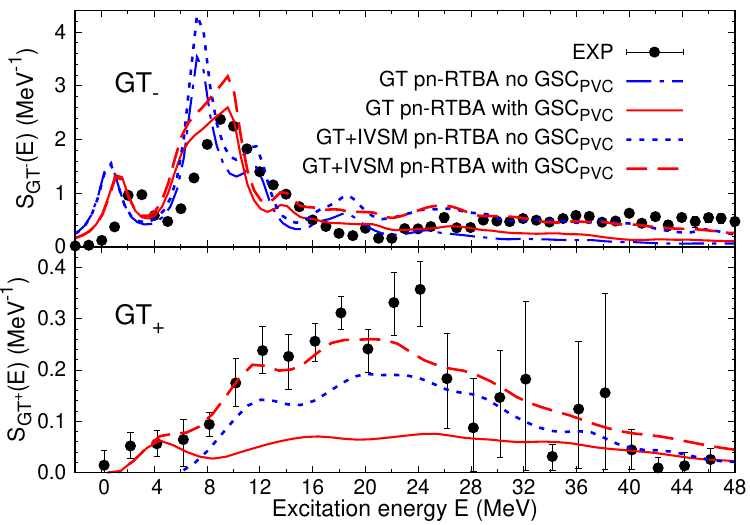}
\end{center}
\caption{GT strength distributions for the transitions $^{90}\mbox{Zr} \rightarrow ^{90}\mbox{Nb}$ (top) and $^{90}\mbox{Zr} \rightarrow ^{90}\mbox{Y}$ (bottom).  The pure GT strength and the mixed GT+IVSM strength of pnRTBA without GSC$_{PVC}$ (dashed-dotted and dotted blue) and with GSC$_{PVC}$ (solid and dashed red) are displayed, in comparison to the experimental data \cite{Yako2005,Wakasa1997}. The figure is adopted from Ref. \cite{Robin2019}.}
\label{f:90Zr}%
\end{figure}

The most advanced calculations with the PVC kernel included, in addition to the standard NFT terms, also the ground state correlations (GSC) caused by PVC (GSC$_{PVC}$). These correlations were introduced and discussed in detail, for instance, in Refs. \cite{KamerdzhievTertychnyiTselyaev1997,KamerdzhievSpethTertychny2004}, where their role in the spin-flip magnetic dipole excitations was found significant. As the GTR also involves the spin-flip process, an important contribution from the GSC-PVC is expected. It can be especially significant in the GT$_+$ branch in neutron-rich nuclei, where these correlations were found to be solely responsible for the unblocking mechanism \cite{Robin2019}. An example is given in Fig. \ref{f:90Zr}, where the GT$_{\pm}$ strength distributions in $^{90}$Zr are shown in comparison to data of Refs. \cite{Yako2005,Wakasa1997}. 
In the GT$_-$ branch of the response, the inclusion of the PVC effects within the pnRTBA leads to an overall fragmentation and broadening of the strength distribution, as compared to the pnRRPA (not shown).
In the GT$_+$ branch, in principle, the GSC of RPA (GSC$_{RPA}$) can unlock transitions from particle to hole states, but such transitions appear only above 7 MeV with very low probabilities.
The inclusion of PVC in the pnRTBA with only the standard NFT forward-going diagrams in the PVC kernel induces almost no change.
However, the inclusion of the GSC$_{PVC}$ associated with backward-going PVC processes has a very strong effect on the GT$_+$ strength. These correlations cause fractional occupancies of the single-particle states of the parent nucleus, which leads to new transitions from particle to particle state and from hole to hole state. For instance, the peak around $4.5$ MeV appears mainly due to the $\pi 1g_{9/2}\to \nu1g_{7/2}$ and $\pi 2p_{3/2} \to \nu 2p_{1/2}$ transitions, with the corresponding absolute values of the transition densities (\ref{trden}) of $0.347$ and $0.182$, respectively.

In the calculations shown in Fig. \ref{f:90Zr}, the theoretical GT$_+$ and GT$_-$ strength distributions were smeared with a parameter $\Delta = 2$ and $1$ MeV, respectively, to match the experimental energy resolutions. 
As in the case of electromagnetic excitations, the pnRRPA calculations do not provide a good agreement with data; therefore, they are not shown. In the GT$_-$ channel, the pnRTBA with GSC$_{PVC}$ demonstrates a good agreement with the data up to $\sim 25$ MeV, except for a small mismatch of the position of the low-lying state.
Remarkably, in the GT$_+$ channel, the GSC induced by PVC are solely responsible for the appearance of both the low-energy peak at $4$ MeV and the higher-energy strength up to $\sim 50$ MeV. Above the low-lying peak, even the pnRTBA GT$_+$ strength alone largely underestimates the data. It is well known, however, that at large excitation energy contributions of the isovector spin-monopole (IVSM) mode become important.  The data of Refs. \cite{Yako2005,Wakasa1997}, in particular, also contain the contribution of the IVSM excitations, which could not be disentangled from the GT transitions due to technical difficulties.
The IVSM modes are generated by response to the operator $F_{IVSM}^{\pm} = \sum_{i} r^2(i) {\overrightarrow\Sigma}(i) \tau_{\pm}(i)$, which should be mixed with the GT response, for instance, following the procedure of Ref. \cite{Terasaki2018}. It introduces the mixed operator $F_{\alpha}^{\pm} = \sum_{i} [1 + \alpha r^2(i)] {\overrightarrow\Sigma}(i) \tau_{\pm}(i)$, where $\alpha$ is a parameter adjusted to reproduce the magnitude of the theoretical low-energy GT strength. 
In this way, the values  $\alpha = 9.1 \times 10^{-3}$ and $\alpha = 7.5 \times10^{-3}$ fm$^{-2}$ were adopted for the GT$_+$ and GT$_-$ branches, respectively.
After that, as one can see from Fig. \ref{f:90Zr}, the resulting strength above 25-30 MeV reasonably describes the data in the (p,n) branch, thus highlighting the importance of both the GSC$_{PVC}$ and the IVSM contribution.
In the (n,p) channel the results are also improved after adding the GSC$_{PVC}$ and the IVSM in pnRTBA, so that  a very good agreement of the overall strength distribution is obtained also for GT$_-$. Further details and discussions of this case can be found in Ref. \cite{Robin2019}.

\begin{figure}[t]
\begin{center}
\includegraphics[width=0.7\textwidth]{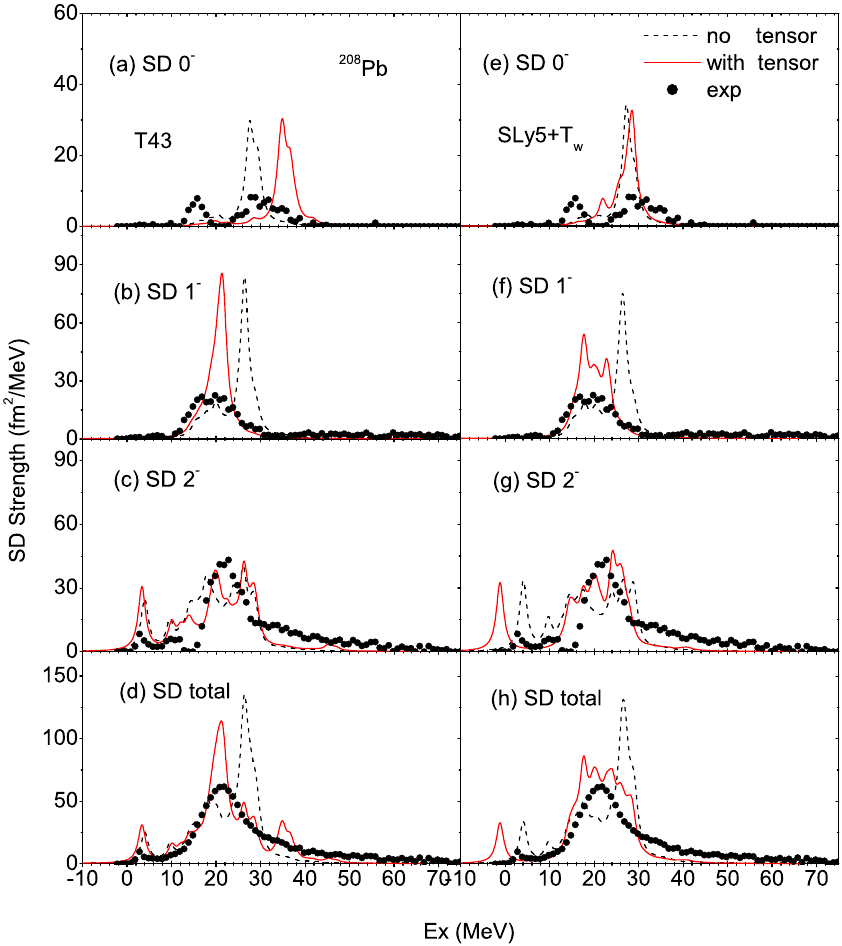}
\end{center}
\caption{Spin-dipole strength distributions in $^{208}$Pb, calculated by Skyrme RPA without and with the tensor force. The discrete RPA results have been smoothed by $\Delta = 2$~MeV and
compared with the experimental data \cite{Wakasa2012}. The figure is taken from Ref.~\cite{Bai2010}.}
\label{Fig:Bai2010}%
\end{figure}

The nucleon-nucleon tensor force is another hot topic in the past two decades \cite{Sagawa2014, Otsuka2020}.
Figure~\ref{Fig:Bai2010} displays the SDR strength distributions in $^{208}$Pb in the $T_-$ channel. Panels (a)--(c) and (e)--(f) show the $J^\pi = 0^{-}, 1^{-}, 2^{-}$ multipoles, respectively, and panels (d) and (h) illustrate the total strength distributions. The experimental data with the multipole decomposition were obtained only recently with the polarized proton beam \cite{Wakasa2012}. It is remarkable that the centroid energies of SDR in $^{208}$Pb are found to be $E(1^-) \approx E(2^-) < E(0^-)$, instead of $E(2^-) < E(1^-) < E(0^-)$ conjectured by the most conventional shell-model and RPA calculations.
The study of Ref.~\cite{Bai2010} analyzed the sensitivity of the SDR centroids to the properties of the tensor force. It is shown that the conventional Skyrme RPA calculations without the tensor force, in general, fail in reproducing the experimentally established relationship between the centroids of the three components of the SDR. Only when the tensor force with specific signs and strengths is included as, e.g., in the effective interactions T43 and SLy5+T$_{\rm W}$, the centroid of the $1^-$ component is significantly pushed down, while that of the $2^-$ component is slightly pushed up. In such a way, the experimental data on the centroids order are reproduced.

In the relativistic framework, to include the tensor force, the Fock terms of the meson-exchange must be taken into account. This is the relativistic Hartree-Fock (RHF) theory \cite{Bouyssy1987, LongVanGiaiMeng2006}. It is, however, not straightforward to identify the tensor effects in the RHF theory, because the tensor force is mixed together with other components, such as the central and spin-orbit ones. For example, simply excluding the pion-nucleon coupling, which is known as the most important carrier of the tensor force, leads to substantial changes also in the central part of the mean field. The quantitative analysis of tensor effects in the RHF theory was achieved for the first time in Ref.~\cite{Wang2018}, which allows fair and direct comparisons with the corresponding results in the non-relativistic framework. It is found that the strengths of tensor force in the existing RHF effective interactions are, in general, weaker than those in the non-relativistic Skyrme and Gogny theories \cite{Wang2018, Wang2020}. So far, the SDR in $^{208}$Pb has not been reproduced within the RHF+RPA scheme yet, with reasonable strengths of the tensor force constrained by the covariant symmetry. The study of this open question is in progress. One of the possible ways is to establish a bridge between the relativistic and non-relativistic DFTs, by performing the non-relativistic expansion with fast convergence \cite{Guo2019PRC, Guo2020, Ren2020}.

\section{\textit{Implications for astrophysics and outlook}}


In this Chapter, the authors discuss the nuclear response theory---the exact equation of motion and its hierarchy of approximations to the response function of an atomic nucleus.
On the one hand, it can be seen from the selected applications shown above that the recent theoretical developments beyond RPA and their numerical implementations have substantially improved the microscopic description of nuclear spectral properties, in particular, compared to the phenomenological models and the conventional RPA methods.
On the other hand, it is also seen that further efforts on advancing the nuclear response theory are needed to obtain an even more accurate description of nuclear spectral properties.

Besides being an interesting theoretical problem, the response theory has many applications, where accurate nuclear excitation spectra are required, especially at the extremes of energy, mass, isospin, and temperature. The most prominent example is nuclear astrophysics, in particular, the rapid neutron capture process ($r$-process) nucleosynthesis in kilonova, core-collapse supernovae, and neutron star mergers \cite{Kajino2019}. The nuclear response to the electric and magnetic dipole, Gamow-Teller and spin-dipole operators are the microscopic sources of the major astrophysical reaction rates, such as the radiative neutron capture 
$(n,\gamma)$, electron capture, $\beta$ decay, and $\beta$-delayed neutron emission. These rates are very sensitive to the fine details of the calculated response, or strength functions, in the given channels, and needed for many nuclei, including those, which are not accessible in laboratory. The low-energy parts of the listed strength distributions are of particular importance. The low-lying dipole strength, which is relevant for the $(n,\gamma)$ rates, was studied very intensively during the past decades and associated with the neutron skin oscillations. In the neutron-rich nuclei, lying on the r-process path in the nuclear landscape, such oscillations form the pygmy dipole resonance, which can affect the $(n,\gamma)$ rates considerably \cite{LitvinovaRingTselyaevEtAl2009,LitvinovaLoensLangankeEtAl2009,
SavranAumannZilges2013,PaarVretenarKhanEtAl2007}. The low-energy parts of the GTR and SDR are responsible for the beta decay and electron capture rates \cite{NiksicMarketinVretenarEtAl2005a,Niu2013,Mustonen2016,Dzhioev2020}.  The recent developments have demonstrated, in particular, that the weak reaction rates are affected considerably by the nuclear correlations beyond (Q)RPA \cite{Niu2013,NiuNiuColoEtAl2015,RobinLitvinova2016,RobinLitvinova2018,LitvinovaRobinWibowo2020,Litvinova2021a}.
Nevertheless, the simplistic (Q)RPA theoretical reaction rates as well as the mean-field nuclear matter equation of state are still employed in most of astrophysical simulations, while the deficiencies of these approaches are even more amplified in stellar environments \cite{ArnouldGorielyTakahashi2007,MumpowerSurmanMcLaughlinEtAl2016,Langanke2021,Cowan2021}. Therefore, adopting the microscopic methods advanced beyond QRPA for astrophysical simulations can be the first step on the way to a high-quality nuclear physics input for such simulations.

The inability of the theory to provide accurate nuclear spectra impedes the progress on other related disciplines, including the searches for the new physics beyond the Standard Model in the nuclear domain, such as the neutrinoless double $\beta$ decay and the electric dipole moment. These applications involve a delicate interplay of numerous emergent effects beyond QRPA and, thus, also require computation of consistency and accuracy, which are beyond the limits of current state-of-the-art theoretical and computational approaches to nuclear response.
A major hope to resolve the issues discussed above is to reconcile consistently the static and dynamical kernels of the EOMs for the nuclear response in various channels, based on the lessons learned from the existing approaches. This has to be complemented by a strong effort on the nuclear interactions, both on the bare and the effective interactions, where the meson-exchange interactions \cite{Machleidt1989}, the chiral effective field theory ($\chi$EFT) \cite{Epelbaum2020,Kolck2020}, and the DFT \cite{Meng2016,Colo2020} are the most promising ones. 


\section{\textit{Acknowledgement}}
This work is supported in part by the US-NSF Career Grant PHY-1654379,
the JSPS Grant-in-Aid for Early-Career Scientists under Grant No.~18K13549,
the JSPS Grant-in-Aid for Scientific Research (S) under Grant No.~20H05648,
RIKEN iTHEMS program,
and RIKEN Pioneering Project: Evolution of Matter in the Universe.

\bibliographystyle{h-physrev}
\bibliography{BibliographyOct2021}

\end{document}